# Manifold Learning with Implicit Physics Embedding for Reduced-Order Flow-Field Modeling

Weiji Wang, Chunlin Gong, Xuyi Jia and Chunna Li[*]

Shaanxi Aerospace Flight Vehicle Design Key Laboratory, School of Astronautics, Northwestern Polytechnical University, Xi'an 710072, People's Republic of China

*Corresponding author. Email address: chunnali@nwpu.edu.cn (C. Li)

**Abstract**

Nonlinear manifold learning (ML) based reduced-order models (ROMs) can substantially improve the quality of nonlinear flow-field modeling. However, noise and the lack of physical information often distort the dimensionality-reduction process, reducing the robustness and accuracy of flow-field prediction. To address this problem, we propose a novel manifold learning ROM with implicit physics embedding (IPE-ML). Starting from data-driven manifold coordinates, we incorporate physical parameters (e.g., angle of attack, Mach number) into manifold coordinates system by minimizing the prediction error of Gaussian process regression (GPR) model, thereby fine-tuning the manifold structure. These adjusted coordinates are then used to construct a flow-fields prediction model that predict nonlinear flow-field more accurately. The method is validated on two test cases: transonic flow-field modeling of the RAE2822 and supersonic flow-field modeling of the hexagon airfoil. The results indicate that the proposed IPE-ML can significantly improve the overall prediction accuracy of nonlinear flow fields. In transonic case, shock-related errors have been notably reduced, while in supersonic case the method can confine errors to small local regions. This study offers a new perspective on embedding physical information into nonlinear ROMs.

## 1. Introduction

Computational fluid dynamics (CFD) technology, the predominant method for solving flow fields, suffers from high computational cost, limiting its use in real-time flow control, surface-load monitoring during aircraft flight, and rapid early-stage design iteration of cars and aircrafts. Therefore, the reduced-order models (ROMs) have emerged as a critical strategy in fluid mechanics, offering a new way to approximate full-order flow fields with much fewer degrees of freedom, thereby enabling flow-field solution to be completed in seconds or even milliseconds.

The most representative and widely-used ROMs are proper orthogonal decomposition (POD), including intrusive POD and nonintrusive POD, and dynamic mode decomposition (DMD). First introduced by Lamely [1], POD reduces the order of flow fields by projecting them onto a low-dimensional linear subspace spanned by a set of orthogonal bases. Subsequently, Galerkin projection or interpolation techniques can be employed to construct intrusive POD or nonintrusive POD models for flow-fields prediction. Siena [2] applied POD-Galerkin to convection-dominated incompressible-flow modeling and introduced two strategies to strengthen the stability and accuracy of the conventional POD-Galerkin method under strong convection conditions. Li [3] employed nonintrusive POD for reduced-order modeling of parametrized time-dependent problems, decoupling spatial and temporal parameters through two-level POD bases. In the work of Hijazi [4], a combination of intrusive POD and nonintrusive POD was proposed, with data-driven nonintrusive POD used to predict eddy viscosity and Galerkin-based intrusive POD employed to solve the velocity and pressure fields. DMD decomposes unsteady flow fields in the frequency domain [5]. Wu [6] utilized DMD assisted with deep-learning method to predict flow from sparse data. Using three distinct interpolation methods together with DMD, Du [7] constructed a parameterized ROM for flow data at target parameters. At the same time, Zhang [8] proposed a mode-selection strategy based on fractal dimension feature embedding to enhance the accuracy and interpretability of traditional DMD. Moreover, several researchers also compared POD and DMD across multiple datasets ([9], [10], [11], [12], [13]). The results show that DMD outperforms POD in extracting modes with specific frequencies, whereas POD achieves smaller overall reconstruction errors on most datasets with fewer modes. Based on this, Jia [14] proposed a hybrid approach: DMD extracts the dominant components of the unsteady flow, while POD reconstructs the residual flow, thereby significantly improving flow-field prediction accuracy.

However, linear-subspace-based ROMs often struggle when facing flows with strong nonlinearities, such as shocks and flow separation. Although POD and DMD have been extended by several researchers—e.g., multi-scale POD ([15]), multi-resolution POD [16], and nonlinear DMD [17]—their inherent linear assumptions restrict their applicability to modeling strongly nonlinear flows. These challenges highlight the need for more expressive ROMs that can capture the intrinsic low-dimensional nonlinear manifold of the flow fields beyond linear subspaces. In recent years, nonlinear dimensional reduction techniques have attracted increasing interest for reduced-order modeling of complex flows. Under the assumption that high-dimensional flow data may lie near a low-dimensional manifold (or latent space), nonlinear ROMs constructed with autoencoder (AE) and manifold-learning (ML) techniques have shown great potential in improving accuracy of flow-field prediction. Raj [18] employed convolutional AE to reduce the dimension of chaotic flow in random arrangement of cylinders and predicted future states with deep learning (DL) methods. It is found that the constructed nonlinear DL models are more accurate than traditional DMD model. Kim [19] developed a physics-informed AE to accelerate the solution of 1-D and 2-D Burgers equations. Additionally, AE has also been applied to the modeling of transonic flows [20], complex vortex [21], and turbulence ([22], [23] [24]). However, the expensive model training makes AE-based ROMs difficult to directly deploy in engineering application. In contrast, ML approaches offer greater advantages in computational efficiency and cost. On the analysis of complex flows, Tauro [25] utilized isometric feature mapping (Isomap) [26] to identify flow patterns of the wake past a circular cylinder from steady to turbulent flows. Farzamnik [27] employed Isomap to analyze and reconstruct several types of shedding-dominated shear flows. Marra [28] observed a correlation between the manifold coordinates and the physical parameters when applying Isomap to reduce the order of controlled flows. Meanwhile, ML–based methods for flow-field modeling have also been substantially developed. In 2014, Franz [29] pioneeringly applied Isomap integrating interpolation algorithm to predict the transonic flows of airfoils and wings. Several researchers further extended the method based on his work. Decker [30] constructed a multi-fidelity nonlinear ROM by manifold alignment and Isomap to reduce the dependence on high-fidelity data. Zheng [31] employed locally linear embedding (LLE) [32] for dimensional reduction and incorporated distance constraints on manifold during interpolation, thereby significantly improving prediction accuracy with a small number of samples. Li [33] successfully predicted full speed flow fields by a ROM combined with Isomap and LLE. Wang [34] proposed a nonlinear flow-field reconstruction method that improves the accuracy of the inverse mapping from manifold coordinates to the full-order flow fields.

  Nevertheless, these ML-based flow-fields modeling methods mentioned above share a common drawback: the manifold coordinates are obtained through data-driven ML, while physical parameters (e.g., angle of attack, Mach number.) are only used for interpolation, thereby limiting the models' generalizability. Moreover, the presence of noise may bias the manifold coordinates obtained by data-driven ML, leading to inaccurate dimensional reduction and consequently reducing prediction accuracy. Therefore, it becomes crucial to reduce dimensionality through nonlinear ROMs, as well as to embed the effect of physical parameters into the reduced-order manifold coordinate system.

To address this problem, we propose a novel nonlinear ROM which implicitly embeds physical parameters into the ML method to enhance the accuracy of flow-field prediction by an iterative refinement of reduced-order manifold. Unlike conventional data-driven ROMs, our method learns a physical-parameter-aware manifold representation which captures parameter-related flow structures and systematically reduces the errors introduced during surrogate-model prediction.

The paper is organized as follows: in section 2, the developed manifold learning with implicit physics embedding (IPE-ML) method is introduced in detail; then the method validation and comparison are carried out and discussed on 2 test cases: transonic flow around RAE2822 and supersonic flow around hexagon airfoil in section 3; In section 4, we draw the conclusions and outlines the future works.

## 2. Manifold learning with implicit physics embedding

In this work, we implicitly embed physical information into low-dimensional through optimizing the initial manifold coordinates obtained by classic data-driven ML methods. The overall procedure is shown in Figure 1, which mainly includes 5 steps:

**Step 1:** Generate samples in the sampling space of physical parameters. The RANS-based CFD simulations are then performed to acquire the full-order flow fields $X$.

**Step 2:** Compute initial low-dimensional manifold coordinates matrix $Y^{(0)}$. This could be achieved by the classic isometric feature mapping (Isomap) or local linear embedding (LLE). In this step, $Y^{(0)}$ is acquired by purely data-driven ways, leading to absence of the physical information.

**Step 3:** Construct initial surrogate model $f^{(0)}$. The surrogate model is used to predict the manifold coordinates by physical parameters $\Theta$ (e.g., $\alpha$, $M_\infty$), thus physical information is implied in the output of the surrogate model implicitly.

**Step 4:** Optimize the manifold coordinates $Y$. The optimization problem of the difference between $Y$ and the surrogate's prediction $\hat{Y}$ is solved by L-BFGS. During this step, we expect that the overall distribution of the samples in the manifold space can be slightly adjusted under the constrains of physical information implied in the output of the surrogate model. Through optimization, an optimized manifold coordinates matrix $Y^*$ and a well-trained surrogate model $f^*$ are obtained.

**Step 5:** Predict new flow fields. The manifold coordinates of new samples $Y_{prediction}$ are predicted through the physical parameters $\Theta_{prediction}$ and $f^*$, then an inverse mapping is performed to reconstruct the high-dimensional flow fields $X_{prediction}$.

The method implementation is as shown in Algorithm 1.

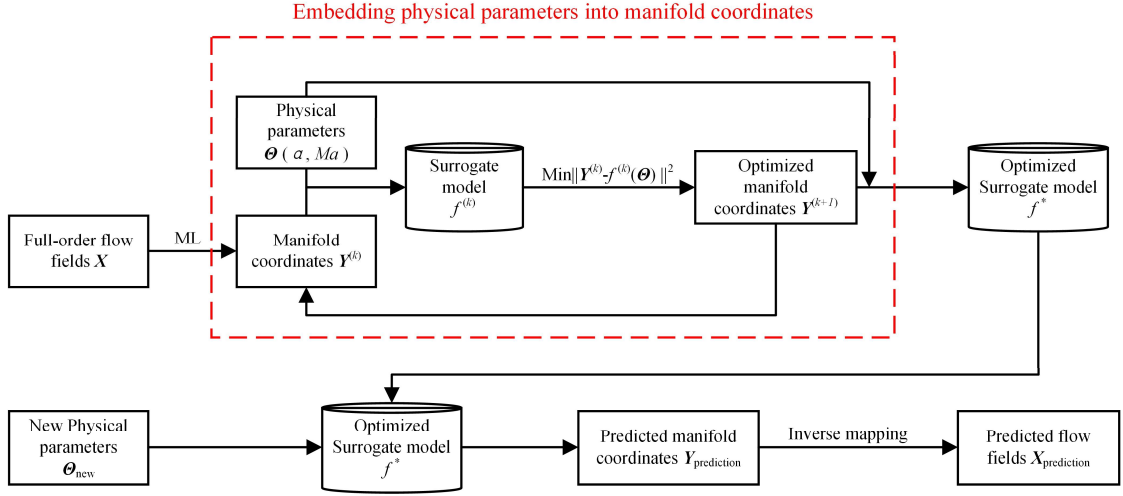

Figure 1 The overall procedure of IPE-ML.

Algorithm 1 Manifold learning with implicit physics embedding (IPE-ML)

---

**Input:** physical parameters space $Q$ (e.g. $\alpha$, $M_\infty$), physical parameters for prediction $\Theta_{prediction} = [\theta_1, \theta_2, ..., \theta_n]^T$.

**Output:** predicted full-order flow fields $X_{prediction} \in \mathbb{R}^{n \times D}$.

1: $\Theta = [\theta_1, \theta_2, ..., \theta_N] \sim Q$, $X = $ RANS-CFD$(\Theta) \in \mathbb{R}^{n \times D}$; // Step 1. Data generation

2: $Y^{(0)} = $ Manifold Leaning$(X) \in \mathbb{R}^{n \times d}$; // Step 2. Initial manifold learning

3: $f^{(0)}: \Theta \to Y^{(0)}$; // Step3. Initial surrogate model construction

4: **while** $loss > \varepsilon$ **and** $k < K$ **do**

5: $\quad loss = \left\| Y^{(k)} - f^{(k)}(\Theta) \right\|_F^2$;

6: $\quad Y^{(k+1)} = \arg\min_Y (loss)$; // optimize manifold coordinates through L-BFGS

7: $\quad f^{(k+1)}: \Theta \to Y^{(k+1)}$; // Update surrogate model

8: $\quad Y^* = Y^{(k+1)}$;

9: $\quad f^* = f^{(k+1)}$;

10: **end while** // Step 4. Manifold coordinates optimization with surrogate model update

11: $Y_{prediction} = f^*(\Theta_{prediction}) \in \mathbb{R}^{m \times d}$;

12:  $X_{\text{prediction}} = \text{InverseMapping}(Y_{\text{prediction}})$; // Step 5. New flow field prediction

## A. Obtaining initial manifold coordinates via manifold learning

To obtain initial manifold coordinates $Y^{(0)}$, we first reduce the dimensionality of the training dataset using traditional manifold learning methods. As our IPE-ML method is general, we employ two representative methods: local linear embedding (LLE) and isometric feature mapping (Isomap). The corresponding IPE-ML methods are named as IPE-LLE and IPE-Isomap, respectively.

According to the classification of Lee [35], the LLE is a topology-preserving approach that aims to keep the linear relationship of neighbor, while the Isomap is a distance-preserving method which means maintaining the geodesic distance between samples in the original space and the manifold.

Assume that $n$ vectorized flow-field samples $x_i \in \mathbb{R}^D$ form a matrix $X = [x_1, x_2, ..., x_D]^T \in \mathbb{R}^{n \times D}$, the basic processes of the Isomap and LLE is introduced as follow.

### a) Local linear embedding

The basic assumption of LLE is that each sample can be reconstructed by a linear combination of its neighbors on both manifold and the original space. First, we find the $k$ nearest neighbors of each sample and obtain their index $Q_i$. The weight $w_{ij}$ of the $i$th sample's $j$th neighbor can be solved by

$$\min_{w_1, w_2, ..., w_n} \sum_{i=1}^{n} \left\| x_i - \sum_{j \in Q_i} w_{ij} x_j \right\|_2^2 \quad (1)$$
$$\text{s.t.} \sum_{j \in Q_i} w_{ij} = 1$$

Formula (1) is often solved by the Lagrange multiplier method. On the manifold, each sample is reconstructed using weights and neighbors which are the same with the original space, so the low-dimensional manifold coordinates $Y = [y_1, y_2, ..., y_n]^T \in \mathbb{R}^{n \times d}$ can be computed by

$$\min_{y_1, y_2, ..., y_n} \sum_{i=1}^{n} \left\| y_i - \sum_{j \in Q_i} w_{ij} y_j \right\|_2^2 = \min_{Y} \| Y - WY \|_F^2 \quad (2)$$

Let

$$M = (I_n - W)(I_n - W)^T \quad (3)$$

then Formula (2) is equivalent to

$$\min_{Y} \mathrm{tr}(YMY^{\mathrm{T}}) \tag{4}$$

Perform eigenvalue decomposition of matrix $M$

$$M = V\Lambda V^{\mathrm{T}} \tag{5}$$

where $V = [v_1, v_2, ..., v_n]^{\mathrm{T}} \in \mathbb{R}^{n \times n}$ is the eigenvector matrix and $\Lambda = \mathrm{diag}(\lambda_1, \lambda_2, ..., \lambda_n)$ ($\lambda_1 < \lambda_2 < ... < \lambda_n$) is the eigenvalue matrix.

The manifold coordinates matrix $Y$ is formed by the eigenvectors corresponding to the $d$ smallest eigenvalues

$$Y = V_d = [v_1, v_2, ..., v_d]^{\mathrm{T}} \tag{6}$$

**b) Isometric feature mapping**

The Isomap can be regarded as a variation of the multiple dimensional scaling (MDS). By using the geodesic instead of the Euclidean distance to measure the distances between samples, the Isomap can efficiently capture the complex and highly nonlinear manifolds.

Similar to LLE, to reduce the dimensionality through the Isomap, the $k$ nearest neighbors of each sample are found, firstly. A neighboring graph $G$ is then constructed. $G$ is a weighted, undirected graph in which the $j$th and $i$th samples are connected by an edge with the weight $dist(i, j)$, where $dist$ is the Euclidean distance. The Floyed-Warshall algorithm is employed to calculate the point-wise shortest distances on $G$ and a shortest distances matrix $D_G$ is obtained. Finally, the classic MDS algorithm is used to solve the low-dimensional manifold coordinates $Y = [y_1, y_2, ..., y_n]^{\mathrm{T}} \in \mathbb{R}^{n \times d}$, where $y_i \in \mathbb{R}^d$ and $d$ refers to the dimension of the manifold. This is achieved by solving the following optimization problem

$$\min_{Y} \left\| YY^{\mathrm{T}} - B \right\|_{\mathrm{F}}^{2} \tag{7}$$

$$B = -\frac{1}{2} H D_G^2 H \tag{8}$$

$$H = I_n - \frac{1}{n} \mathbf{1}\mathbf{1}^{\mathrm{T}} \tag{9}$$

where $\|\cdot\|_{\mathrm{F}}$ is the Frobenius norm and $(\cdot)^2$ is the element-wise product; $I_n \in \mathbb{R}^{n \times n}$ is the identity matrix and $\mathbf{1}$ refers to the all-ones vector.

To solved Formula (7), perform eigenvalue decomposition of matrix $B$

$$B = V\Lambda V^{\mathrm{T}} \tag{10}$$

where $V = [v_1, v_2, ..., v_n]^{\mathrm{T}} \in \mathbb{R}^{n \times n}$ is the eigenvector matrix and $\Lambda = \mathrm{diag}(\lambda_1, \lambda_2, ..., \lambda_n)$ ($\lambda_1 > \lambda_2 > ... > \lambda_n$) is the eigenvalue matrix.

Then select the $p$ largest eigenvalues and their corresponding eigenvectors to

compute the low-dimensional manifold coordinates

$$Y = V_d \Lambda_d^{1/2} \tag{11}$$

**B. Manifold coordinates refinement under physical constraints**

As the initial manifold coordinates $Y^{(0)}$ have been obtained purely by data-driven manifold learning methods, refinement is needed by embedding information of physical parameters. The main procedure is as follows.

We first establish a Gaussian process regression (GPR) surrogate model $f_{\text{GPR}}^{(0)}$ with physical parameters $\Theta$ as input and the initial manifold coordinates $Y^{(0)}$ as output

$$Y^{(0)} = f_{\text{GPR}}^{(0)}(\Theta) \tag{12}$$

Then we employ $f_{\text{GPR}}^{(0)}$ to predict the manifold coordinates of the training dataset, yielding predicted manifold coordinates $\hat{Y}$. To minimize the difference between the manifold coordinates predicted by surrogate model $\hat{Y}$ and those of the training set, the following optimization needs to be performed

$$\min_Y \left\| Y - \hat{Y} \right\|_F^2 \tag{13}$$

Formula (13) demonstrates that the distribution of the optimized manifold coordinates of the training set are consistent with the output distribution of the surrogate model. Since the surrogate model inherently incorporates physical information from the physical parameters; adjusting the manifold coordinates of the training set through Formula (13) enables the implicitly embedding of this information.

The optimization problem is solved by L-BFGS. Since the gradient of the objective function is known

$$\nabla = 2(Y - \hat{Y}) \tag{14}$$

the problem can be solved easily. However, $\hat{Y}$ must be predicted by an updated surrogate model in each iteration, which requires numerous calls to construct the surrogate model. Therefore, we choose the GPR model as the surrogate model.

After optimization, an optimized manifold coordinates matrix of the training set $Y^*$ and a well-trained GPR model $f_{\text{GPR}}^*$ are obtained. $f_{\text{GPR}}^*$ can be used to predict the manifold coordinates of flow fields $Y_{\text{prediction}}$ under new physical parameters $\Theta_{\text{prediction}}$

$$Y_{\text{prediction}} = f_{\text{GPR}}^*(\Theta_{\text{prediction}}) \tag{15}$$

## C. Inverse mapping for flow-field reconstruction

We adopt the inverse mapping method proposed by Wang [34] to reconstruct the high-dimensional flow fields $X_{prediction}$ by $Y_{prediction}$.

Assume a set of modes $W$ that can reconstruct the flow field $X \in \mathbb{R}^{n \times D}$ nonlinearly in kernel space

$$X \approx X' = \Phi(Y, Y)W^T = ZW^T \tag{16}$$

where $\Phi(\cdot, \cdot)$ is the kernel function calculated by 2 matrices; $Y$ is the manifold coordinates of the training set; $Z$ is the kernel matrix.

A ridge regression is employed to solve $W$ by solving the following minimization problem

$$\min_{W}(\|X - ZW^T\|_F^2 + \lambda \|W\|_F^2) \tag{17}$$

where $\lambda$ is the L2 regularization coefficient which is used to prevent overfitting.

Set formula (17) to zero, and the expression of $W$ can be obtained

$$W = X^T Z (Z^T Z + \lambda I_n)^+ \in \mathbb{R}^{D \times n} \tag{18}$$

where $I_n \in \mathbb{R}^{n \times n}$ is the identity matrix and $(\cdot)^+$ represents pseudo-inverse.

For the predicted manifold coordinates $Y_{prediction}$ in formula (15), the predicted flow fields can be reconstructed by

$$X_{prediction} = Z_{new} W^T = \Phi(Y_{prediction}, Y) W^T \tag{19}$$

Different from the previous work, we choose a Matérn kernel in formula (16) to calculate the kernel matrix $Z$. The Matérn kernel is a generalization of the RBF kernel. The kernel value calculated by 2 vectors is given by

$$K(\cdot, \cdot) = \frac{1}{\Gamma(v) 2^{v-1}} \left(\frac{\sqrt{2v}}{l} d(\cdot, \cdot)\right)^v K_v\left(\frac{\sqrt{2v}}{l} d(\cdot, \cdot)\right) \tag{20}$$

where $v$ controls the smoothness and $l$ controls the range of the function; $d(\cdot, \cdot)$ is the Euclidean distance between vectors; $K_v(\cdot)$ is a modified Bessel function; $\Gamma(\cdot)$ the gamma function. When $v \to \infty$, the Matérn kernel is equivalent to the RBF kernel.

## 3. Error analysis

In this section, we first analysis the Lipschitz continuity of the inverse mapping method (KRR-DCR). Then we decompose the sources of prediction error in the proposed IPE-ML method and compare it with the conventional ML-based method (ML+GPR). At last, the range of applicability of our IPE-ML method is presented theoretically.

## A. Lipschitz continuity of inverse mapping

For a single manifold-coordinate vector $y$, define the kernel vector $k(y)$

$$k(y) = \begin{bmatrix} K(y, y_1) \\ \vdots \\ K(y, y_n) \end{bmatrix} \in \mathbb{R}^n \quad (21)$$

According to Formula (19), the reconstructed flow field vector $x$ corresponding to $y$ is

$$x(y) = Wk(y) \in \mathbb{R}^D \quad (22)$$

Thus, for two manifold points $y_1$ and $y_2$, we have

$$\begin{aligned} \|x(y_1) - x(y_2)\|_2 &= \|W(k(y_1) - k(y_2))\|_2 \\ &\leq \|W\|_2 \|k(y_1) - k(y_2)\|_2 \end{aligned} \quad (23)$$

And according to Formula (18), the 2-norm of $W$ is bounded by

$$\begin{aligned} \|W\|_2 &= \|X^\mathrm{T} Z(Z^\mathrm{T} Z + \lambda I_n)^+\|_2 \\ &\leq \|X\|_2 \|Z\|_2 \|(Z^\mathrm{T} Z + \lambda I_n)^+\|_2 \\ &\leq \frac{\|X\|_2 \|Z\|_2}{\lambda} \end{aligned} \quad (24)$$

Furthermore, assume that the selected kernel function is locally Lipschitz continuous, it holds that

$$\|k(y_1) - k(y_2)\|_2 \leq L_k \|y_1 - y_2\|_2 \quad (25)$$

where $L_k$ is the Lipschitz constant for the kernel function. We have

$$\|x(y_1) - x(y_2)\|_2 \leq L\|y_1 - y_2\|_2 + \delta = L_k \frac{\|X\|_2 \|Z\|_2}{\lambda} \|y_1 - y_2\|_2 + \delta \quad (26)$$

where $\delta$ denotes numerical errors and modes-selection errors that arise during the computation.

This result indicates that, when reconstructing flow fields with the KRR-DCR method, deviations in the manifold coordinates are amplified by a finite factor. Consequently, if IPE-ML can reduce the error in the manifold coordinates, the errors in the flow-field reconstruction will also be reduced.

## B. Error decomposition for ML+GPR and IPE-ML

Since KRR-DCR method is Lipschitz continuous on the sampling space, for

ML+GPR, the flow-field prediction error $E_{ML}$ can be expressed as

$$E_{ML} \leq L(E_1 + E_2) + \delta \tag{27}$$

where $E_1$ represents the error of manifold coordinates introduced during dimensionality reduction, attributable to data noise, method feature and an inappropriate selection of neighbors; $E_2$ refers to the prediction error of GPR model. According to the study of Koepernik [36], under appropriate conditions on the kernel and sampling design, the posterior mean of GPR prediction converges to the true function as the number of observations increases, that is $E_2 \to 0$.

Similarly, the prediction error of IPE-ML $E_{IPE-ML}$ can be also decomposed into

$$E_{IPE-ML} \leq L(\tilde{E}_1 + E_2 + B_{bias} + \varepsilon_{opt}) + \delta \tag{28}$$

where $\tilde{E}_1$ is the error between the manifold coordinates optimized by IPE-ML and the true manifold coordinates. In an ideal scenario, after incorporating physical information through the IPE-ML method, $\tilde{E}_1$ should be markedly smaller than $E_1$.

Compared to ML+GPR, the error expression for the IPE-ML includes two additional terms: $B_{bias}$ and $\varepsilon_{opt}$. $B_{bias}$ represents the bias of manifold coordinates during the optimization. When the initial manifold coordinate errors are too large, the optimization has difficulty correcting it and may even adjust the coordinates in a wrong direction, resulting in a larger $B_{bias}$. And $\varepsilon_{opt}$ denotes the residual arising from non-convergence of the optimization. Assuming that, under appropriate settings of the L-BFGS algorithm, the residual of the optimization is sufficiently small and the optimization converges fully, $\varepsilon_{opt} \to 0$.

Since the flow-field reconstruction method and the type of surrogate model are the same between the ML+GPR and IPE-ML, the difference in prediction accuracy mainly stems from errors in the manifold coordinates, that is the difference between $E_1$ and $\tilde{E}_1 + B_{bias}$. It can be further divided into two cases:

(1) $E_1 - \tilde{E}_1 < B_{bias}$

When the number of training samples is insufficient or the ML hyperparameters (number of neighbors, dimensionality of the manifold, et al) are inappropriate, the initial manifold extracted by ML methods may deviate significantly from the true manifold. Consequently, the initial manifold fails to capture the essential characteristics of the flow fields. Adjusting the coordinates using the information of physical parameters based this erroneous manifold tends to shift the distribution of manifold coordinates away from the flow-field features and toward the distribution of the physical parameters. This introduces a significant $B_{bias}$, which is then amplified by the

KRR-DCR algorithm, ultimately producing larger flow-field prediction errors.

(2) $E_1 - \tilde{E}_1 > B_{bias}$

Conversely, when the sample size and hyperparameters are appropriately chosen, the initial manifold already encodes a substantial amount of flow-field features. Incorporating the information of physical-parameters can markedly enhance the physical interpretability and expressive capacity of the manifold coordinates. The improvement in the manifold coordinates ($E_1 - \tilde{E}_1$) achieved by the IPE-ML far exceeds the coordinate bias $B_{bias}$ caused by optimization alone, thereby significantly increasing flow-field prediction accuracy.

## 4. Method validation

In this section, the RAE2822 and hexagon airfoils are presented to validate the ability of our proposed IPE-ML method for transonic and supersonic flow-field prediction under varying angle of attack and Mach number, respectively. Meanwhile, we also compare IPE-ML with traditional data-driven ROMs, including ML and POD, to show the improvement of prediction accuracy after embedding physical information. Across all ROMs mentioned above, the same GPR model is used as the surrogate model to eliminates the influence of surrogate model. The comparative methods are denoted as ML+GPR and POD+GPR, respectively. And among them, ML+GPR built on LLE and Isomap are denoted as LLE+GPR and Isomap+GPR.

The prediction accuracy is qualified by mean relative error $E_{mean}$ and max relative error $E_{max}$, which are respectively defined as

$$E_{mean}(\boldsymbol{x}, \hat{\boldsymbol{x}}) = \operatorname*{mean}_{i \in m}\left(\left|\frac{x_i - \hat{x}_i}{x_i}\right|\right) \tag{29}$$

$$E_{max}(\boldsymbol{x}, \hat{\boldsymbol{x}}) = \operatorname*{max}_{i \in m}\left(\left|\frac{x_i - \hat{x}_i}{x_i}\right|\right) \tag{30}$$

where $\boldsymbol{x}$ is the CFD solution; $\hat{\boldsymbol{x}}$ is the predicted solution; $i$ represents the index of grid point and $m$ refers to the total number of grid points. $E_{mean}$ represents the overall prediction accuracy of the flow fields, while $E_{max}$ measures the accuracy in local regions.

All experiments below are performed on a platform based on the Apple Silicon architecture (M4 processor, 24GB RAM). The LLE, Isomap and GPR are implemented using the scikit-learn package [37] and the L-BFGS is performed using the SciPy package [38].

## A. Test case 1: Transonic flow fields around RAE2822 airfoil

### 1) Dataset generation

The dataset is generated under $M_\infty \in [0.734, 0.85]$ and $\alpha \in [2°, 6°]$. The random Latin hypercube sampling (LHS) is employed to get 550 sample points in the parameter space, in which 500 samples are selected randomly as the training set while the rest 50 as the testing set. Their distribution in the $M_\infty - \alpha$ space is shown in Figure 2.

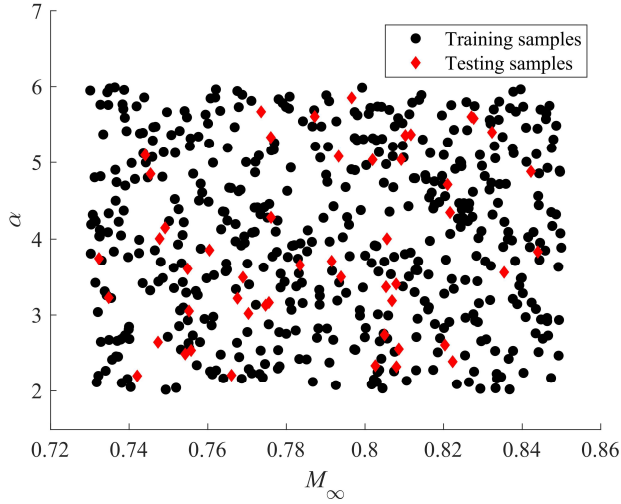

Figure 2 The distribution of transonic samples in parameter space.

We perform RANS-based CFD simulation to obtain the full-order flow fields of the training set. The structured grid is shown in Figure 3, with near-wall resolution maintaining at $y^+ < 1$ and the CFD solver settings is consistent with Ref [34]. The grid convergence study is presented in Appendix A. A density-based implicit solver with SST $k$-$\omega$ turbulence model is utilized. Pressure-far-field boundary condition is applied at the freestream boundaries, while adiabatic, no-slip boundary conditions are imposed on wall surface. The incoming flow is considered as ideal gas. Second-order upwind scheme is selected for spatial discretization, and Roe scheme is employed for flux computation.

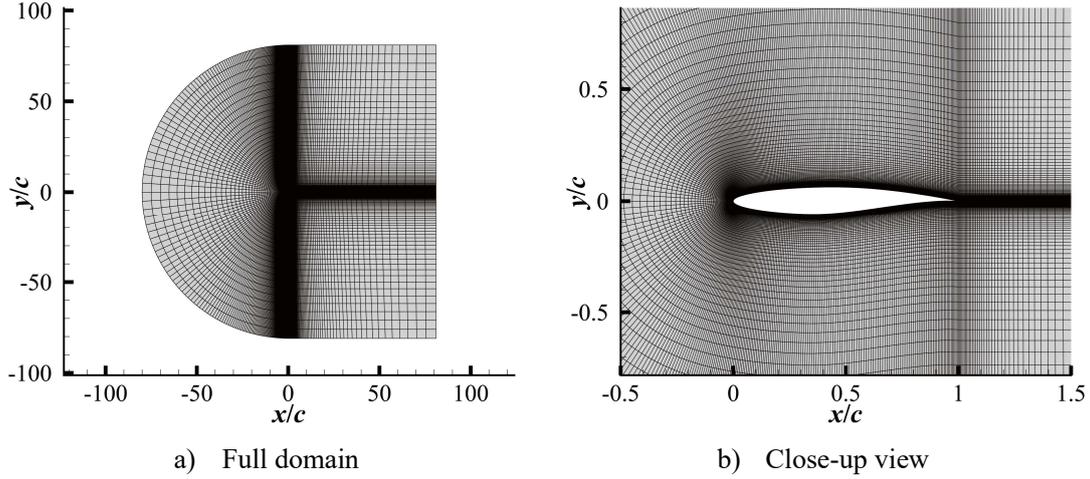

a) Full domain  b) Close-up view

Figure 3 The structured grid for RAE2822 airfoil (82,844 cells in total).

**2) Hyperparameter tuning**

The hyperparameters for the entire process of flow field prediction through proposed IPE-ML are listed in Table 1. During inverse mapping, the L2 regularization parameter $\lambda$ needs to be set as well. Based on the analysis in Ref [34], the impact of $\lambda$ within a small range is smaller than the kernel function parameters on reconstruction accuracy. Therefore, we do not analyze the influence of $\lambda$ on flow-field prediction accuracy in this work, and $\lambda$ is fixed at $1\times 10^{-5}$.

Additionally, the hyperparameters of L-BFGS are not considered in this work as well. The maximum iterations for the optimization is set to 100, and the convergence criteria are the changes in the objective function or the gradient is less than $1\times 10^{-6}$.

Table 1 Hyperparameters during flow-field prediction through IPE-ML.

| Stage | Hyperparameter | Symbol |
|---|---|---|
| Dimensionality reduction | Dimensionality of the manifold | $d$ |
|  | Number of neighbors | $k$ |
| Inverse mapping | Smoothness factor | $v$ |
|  | Range of kernel function | $l$ |

Franz[29] observes that when using Isomap plus interpolation (Isomap+I) for flow-field prediction, the best dimensionality of the manifold $d$ equals to the number of parameters in the design space. Zheng [31] and Li [33] also adopt similar conclusions in their respective works. This is because $d$ represents the number of factors dominating the changes of flow fields, namely the number of design variables.

The number of neighbors $k$ influences the nonlinearity in the dimensionality reduction. A smaller $k$ results in smaller local linear regions and a more nonlinear manifold structure. However, when $k$ is too small, it leads to disconnected neighborhood graphs and failure of dimensionality reduction.

The influence of $k$ on the dimensionality reduction can be assessed through reconstruction error for LLE [39] and residual variance for Isomap [40]. But in order to study its impact on flow-field prediction accuracy more directly, we calculate Formula (21) and Formula (22), then use these results to select the most suitable $k$.

The prediction errors of IPE-ML with dimensionality reduction through LLE (IPE-LLE) for different $d$ and $k$ are shown in Figure 4. The locations with the minimum $E_{\text{mean}}$ and $E_{\text{max}}$ are also marked with red points. It can be seen that the minimums of both $E_{\text{mean}}$ and $E_{\text{max}}$ are located at $d = 2$, which is consistent with the conclusions of existing research. However, the corresponding $k$ values exhibit significantly different influence. A larger $k$ reduces the nonlinearity of the manifold structure, which is beneficial for improving the prediction accuracy in weak nonlinear regions of the flow fields but has a negative impact on the prediction of strong nonlinear regions.

To balance $E_{\text{mean}}$ and $E_{\text{max}}$, we construct a normalized composite error $E_{\text{com}}$

$$E_{\text{com}} = \frac{1}{2}\text{norm}(E_{\text{mean}}) + \frac{1}{2}\text{norm}(E_{\text{max}}) \tag{31}$$

where $\text{norm}(E)$ refers to normalization, which is performed as

$$\text{norm}(E_i) = \frac{E_i - \min(E)}{\max(E) - \min(E)} \tag{32}$$

In Figure 5, $E_{\text{com}}$ for IPE-LLE is the smallest at $d = 2$ and $k = 15$, where the corresponding $E_{\text{mean}}$ and $E_{\text{max}}$ are 0.000402 and 0.1662.

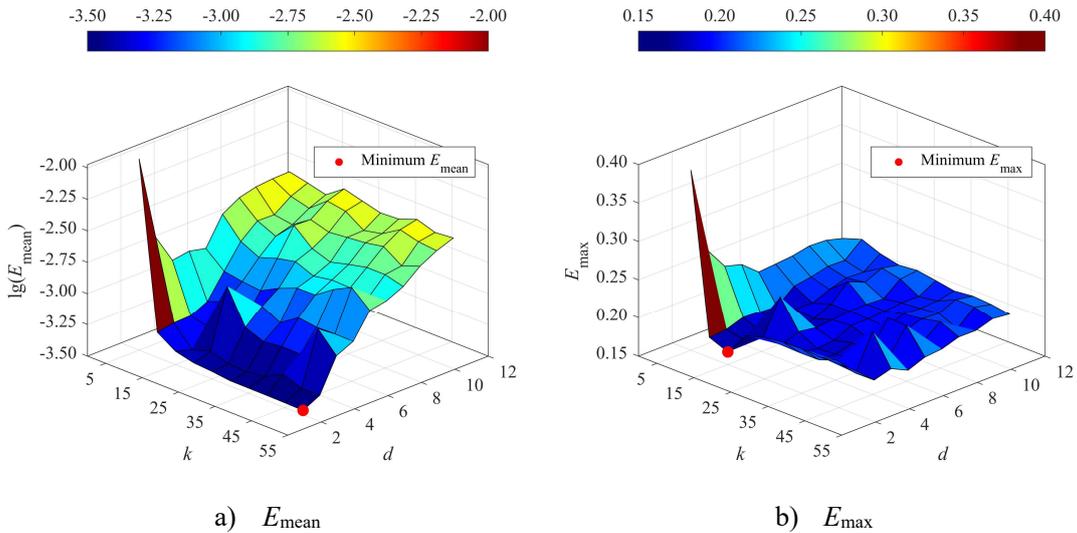

a) $E_{\text{mean}}$        b) $E_{\text{max}}$

Figure 4 The prediction errors of IPE-LLE under different $d$ and $k$.

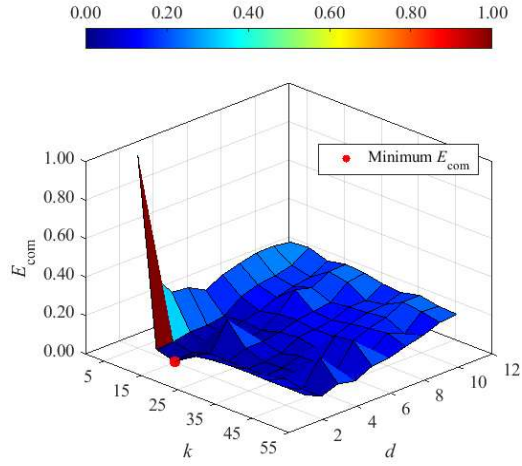

Figure 5 $E_{com}$ of IPE-LLE under different $d$ and $k$.

Similarly, for IPE-ML with dimensionality reduction through Isomap (IPE-Isomap), the variations of $E_{mean}$, $E_{max}$ and $E_{com}$ with $d$ and $k$ are shown in Figure 6 and Figure 7. $E_{com}$ for IPE-Isomap is the smallest at $d = 2$ and $k = 5$, where the corresponding $E_{mean}$ and $E_{max}$ are 0.000378 and 0.1997.

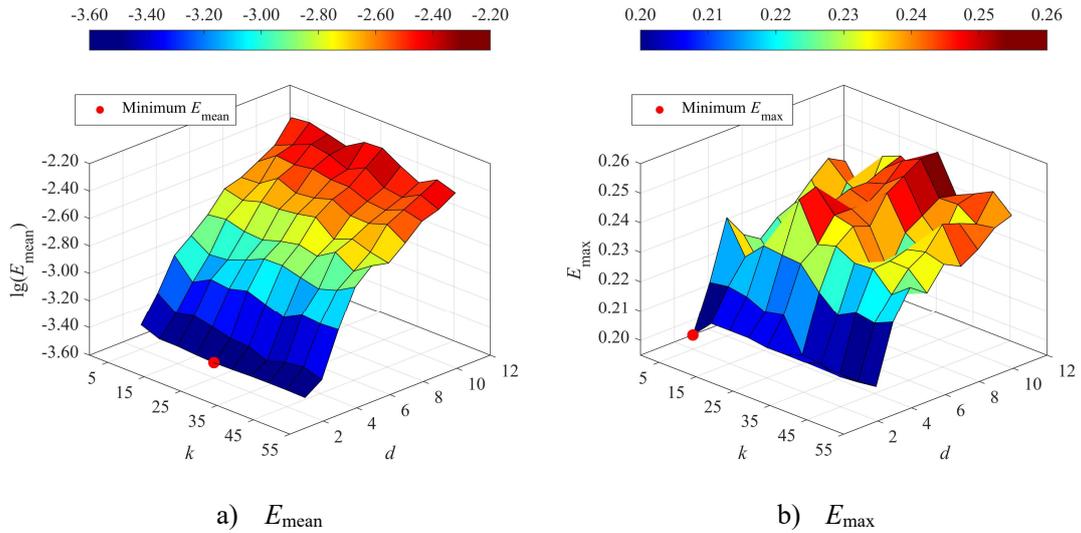

a)  $E_{mean}$  
b)  $E_{max}$

Figure 6 The prediction errors of IPE-Isomap under different $d$ and $k$.

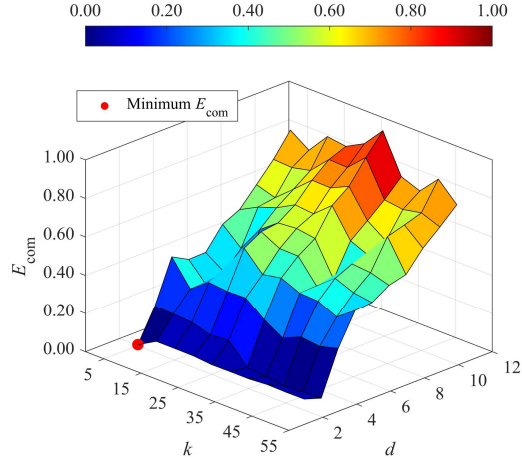

Figure 7 $E_{com}$ of IPE-Isomap under different $d$ and $k$.

The influence of hyperparameters, $v$ and $l$, in inverse mapping accuracy is analyzed then. $E_{mean}$ and $E_{max}$ for the IPE-LLE under different $v$ and $l$ are shown in Figure 8. Both $E_{mean}$ and $E_{max}$ are the smallest near $v = 1.25\text{-}1.5$. Around this value, the kernel function is approximately a first-order differentiable function, and the weights of the samples participating in the flow-field reconstruction within the neighborhood transition more smoothly.

The variation trends of $E_{mean}$ and $E_{max}$ with respect to $l$ are opposite. As the range of the kernel function increases, more global flow-field information is introduced, leading to an increase in the overall reconstruction accuracy. However, excessive global information can lead to significant deviations in local regions (such as regions near shocks). Conversely, decreasing the range of the kernel function results in samples from a small neighborhood participating in the flow-field reconstruction. Since these samples have higher similarity in shock positions and intensities to the flow fields being reconstructed, the local reconstruction accuracy near the shocks can be improved.

$E_{com}$ is also shown in Figure 9. We select the point where $E_{com}$ is the smallest as the optimal $v$ and $l$ ($v = 1.25$ and $l = 0.25$). At this location, $E_{mean} = 0.0004286$ and $E_{max} = 0.1644$.

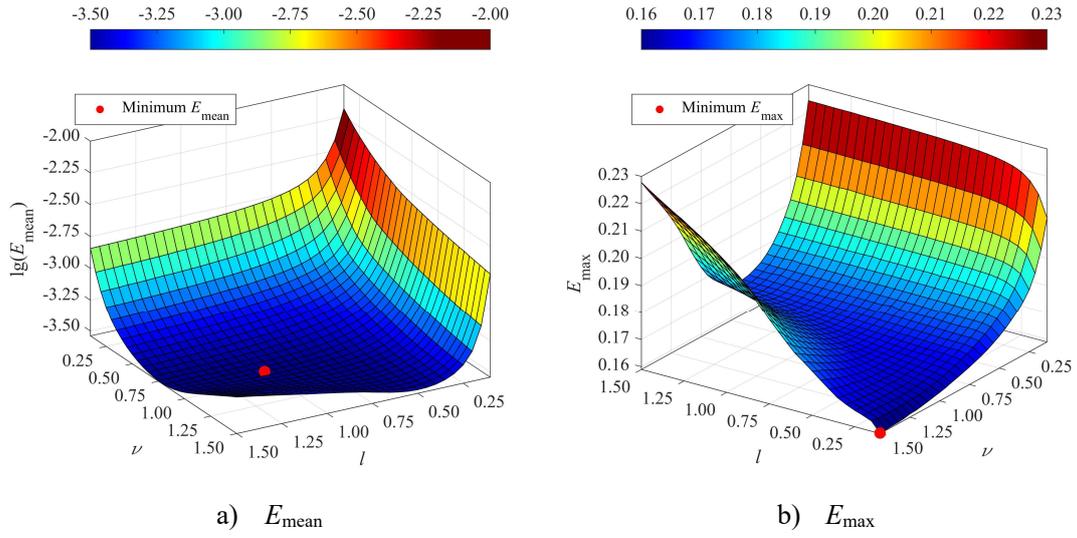

a) $E_{mean}$     b) $E_{max}$

Figure 8 $E_{mean}$ and $E_{max}$ for the IPE-LLE under different $v$ and $l$

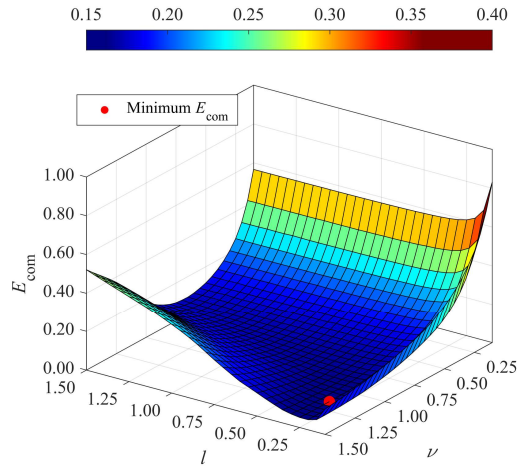

Figure 9 $E_{com}$ under different $v$ and $l$.

Since the IPE-LLE and IPE-Isomap employ the same inverse mapping method, the influence of $v$ and $l$ on the prediction accuracy of IPE-Isomap is not analyzed separately. In conclusion, Table 2 lists the settings of all hyperparameters for the IPE-LLE and IPE-Isomap.

Table 2 Hyperparameters set for the IPE-LLE and IPE-Isomap.

| Method | Hyperparameter | Value |
| --- | --- | --- |
| IPE-LLE | $d$ | 2 |
|  | $k$ | 15 |
|  | $v$ | 1.25 |
|  | $l$ | 0.25 |
|  | *d* | 2 |
|  | *k* | 5 |
| IPE-Isomap | *v* | 1.25 |
|  | *l* | 0.25 |

**3) Flow-field prediction**

After solving the initial manifold coordinates through ML, the convergence history of the optimization for IPE-LLE and IPE-Isomap are shown in Figure 10 and Figure 11, respectively. Both methods reach the convergence requirements within 10 iterations. Figure 12 and Figure 13 compare the original manifold coordinates obtained by the LLE and Isomap with the optimized manifold coordinates obtained by IPE-LLE and IPE-Isomap. The overall shape of the manifold does not change significantly, but adjustments are made locally by embedding physical information. Additionally, for the Isomap, due to a small number of neighbors ($k$=5) and a disconnected neighborhood graph, the manifold structure exhibits some discontinuities. However, IPE-Isomap successfully repairs this problem under the guidance of physical information and gets a more fine-tuned manifold.

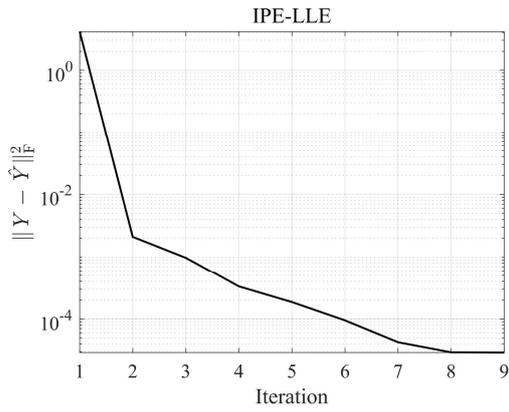
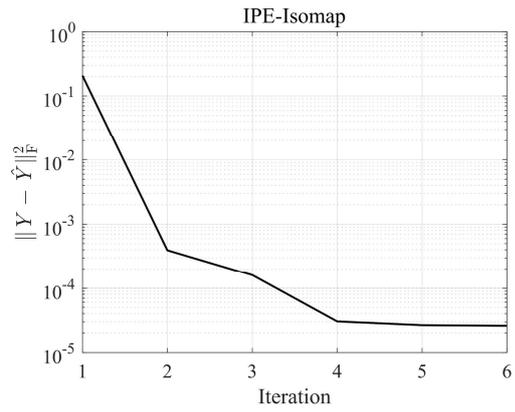

Figure 10 The convergence history for IPE-LLE.

Figure 11 The convergence history for IPE-Isomap.

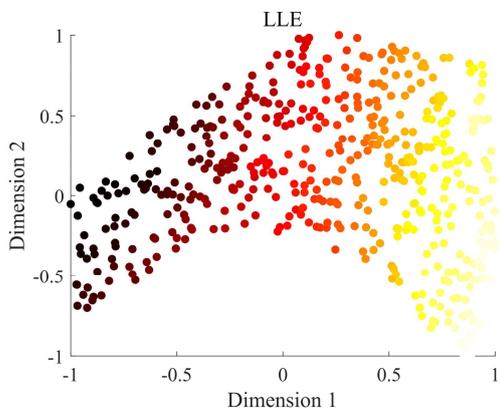
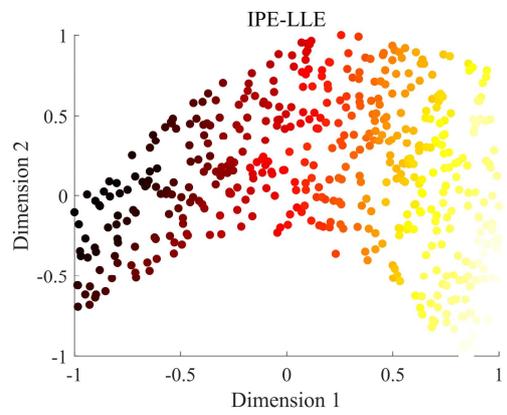

a)  Original LLE coordinates.

b)  Optimized IPE-LLE coordinates.

Figure 12 Comparison of the original manifold coordinates and optimized manifold coordinates for IPE-LLE.

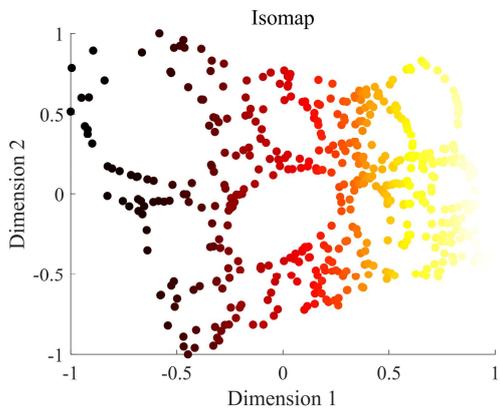
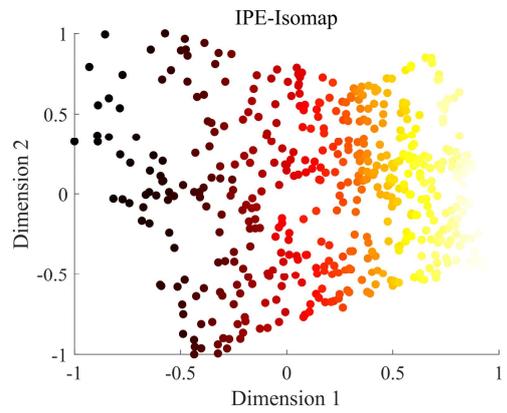

a)  Original Isomap coordinates.

b)  Optimized Isomap coordinates.

Figure 13 Comparison of the original manifold coordinates and optimized manifold coordinates for IPE-Isomap.

Next, we compare the flow-field prediction accuracy of different methods in the light of the error analysis presented in Section 3. The selection of the order POD modes is given in Appendix B.

The initial training sample size and number of neighbors ($k$) influence the structure of the initial manifold. Figure 14 and Figure 15 present a comparison of $E_{mean}$ and $E_{max}$ for the five methods under different sample sizes and $k$. When the training sample size is small, ML methods cannot recover an accurate initial manifold, causing ML-based methods exhibiting even lower flow-field prediction accuracy than POD+GPR. Meanwhile, the adjustments of manifold coordinates by IPE-ML introduces a significant $B_{bias}$, resulting in $E_1 - \tilde{E}_1 < B_{bias}$ and consequently a larger prediction error for IPE-ML relative to ML+GPR. As the number of training samples increases, the initial manifold obtained by ML gradually approaches the true manifold. Therefore, the impact of $B_{bias}$ on prediction accuracy diminishes, and IPE-ML can adjust the obtained manifold toward the true manifold ($E_1 - \tilde{E}_1 > B_{bias}$). In this regime, the flow-field prediction accuracy of IPE-ML is consistently superior to that of ML+GPR.

However, since the LLE's extraction of the manifold relies on local neighborhood, a small $k$ prevents LLE from recovering an appropriate initial manifold regardless of the sample size, resulting in a large $E_1$. Although the IPE-LLE can partially correct the manifold structure, it still cannot guarantee accurate flow-field predictions, as is shown in Figure 15 b).

The results above demonstrate that the applicability of the IPE-ML depends on the ML methods being able to obtain a reasonably accurate initial manifold when provided with a sufficient number of training samples and an appropriate choice of neighbors.

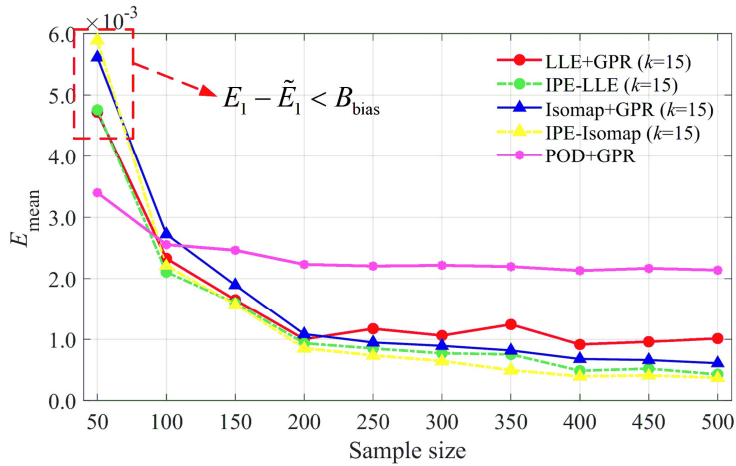

a) $E_{mean}$

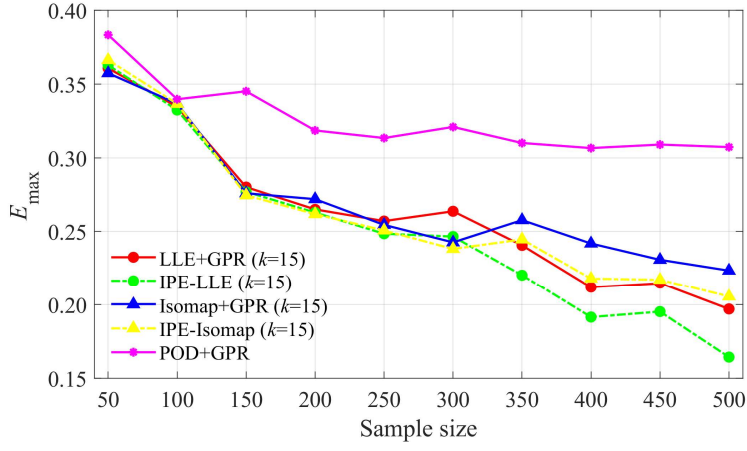

b) $E_{max}$

Figure 14 The flow-field prediction error of 5 different methods under different sample size ($k$=15)

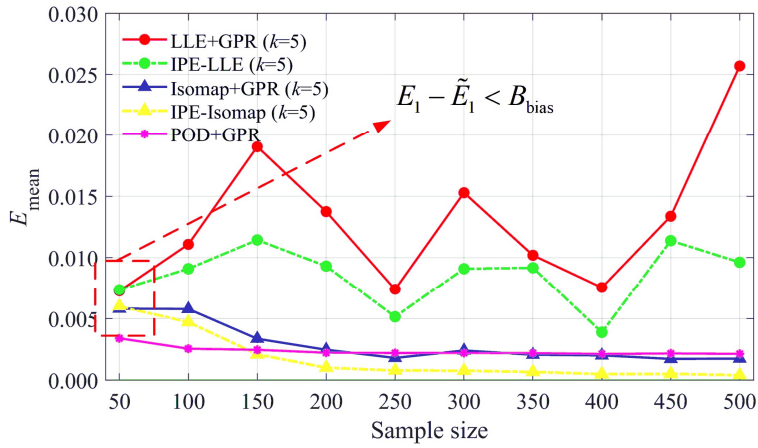

c) $E_{mean}$

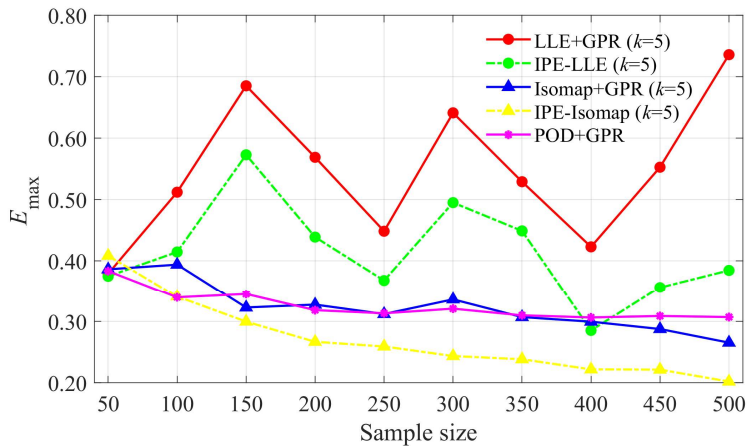

d) $E_{max}$

Figure 15 The flow-field prediction error of 5 different methods under different sample size ($k$=5)

Although $E_{mean}$ for IPE-LLE in Figure 14 and for IPE-Isomap in Figure 15 stabilize when training sample size exceeds 400, $E_{max}$ still decreases markedly. Therefore, we employ the full set of 500 training samples for model construction in the subsequent comparison. Meanwhile, to illustrate the improvement of the IPE-ML over the original ML+GPR methods, the 4 hyperparameters of both LLE+GPR and Isomap+GPR are kept identical to those of their respective IPE-ML.

The $E_{\text{mean}}$ and $E_{\text{max}}$ of the testing set through different methods are compared in Figure 16. The flow-field prediction accuracy of all ML-based methods is significantly higher than that of the POD-based method, demonstrating a better ability of nonlinear ROMs for predicting flow fields with strong nonlinearities such as shocks.

The proposed IPE-ML also shows significant advantages over conventional ML+GPR methods. Since LLE focuses only on the local manifold structure within the neighborhood and ignores the global information, its prediction results exhibit a large global error ($E_{\text{mean}}$). But IPE-LLE effectively address this issue after embedding global physical information into pre-obtained manifold coordinates. Since $M_\infty$ and $\alpha$ dominate the overall evolution of the flow fields, their incorporation can provide the manifold with sufficient global structure information, further massively improving the prediction accuracy.

The discontinuities in the manifold extracted by Isomap+GPR results in much lower prediction accuracy. The manifold fails to reflect the true distances between different samples, further causing significant prediction and reconstruction deviations in the GPR and KRR-DCR models. However, because these distances are related to the variations in $M_\infty$ and $\alpha$, the IPE-Isomap can be used to repair the manifold structure obtained by Isomap. This repairment attributes to the $E_{\text{mean}}$ of IPE-Isomap is only one-fourth of the Isomap+GPR, and $E_{\text{max}}$ is also reduced by more than 23%.

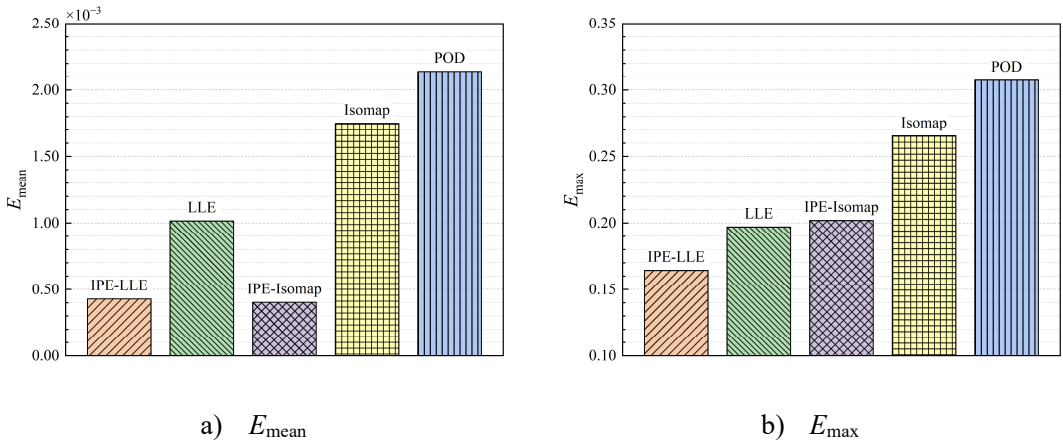

a)  $E_{\text{mean}}$ \qquad b)  $E_{\text{max}}$

Figure 16 $E_{\text{mean}}$ and $E_{\text{max}}$ on the testing set for the different methods.

Furthermore, we randomly select 4 samples from testing set to further compare the abilities to predict flow fields by different methods. The bar chart of the prediction error for the 4 samples is shown in Figure 17. Though LLE+GPR exhibits a smaller $E_{\max}$ than IPE-LLE on the 3$^{\text{rd}}$ sample ($M_\infty = 0.821, \alpha = 4.54°$), it experiences an anomalously large $E_{\text{mean}}$ on the 4$^{\text{th}}$ sample ($M_\infty = 0.830, \alpha = 2.49°$), even higher than POD+GPR. At the same time, the proposed IPE-ML method outperforms POD+GPR on all 4 samples, and for most samples both $E_{\text{mean}}$ and $E_{\max}$ are significantly lower than those achieved by ML+GPR, demonstrating the better generalization capability of the proposed IPE-ML method.

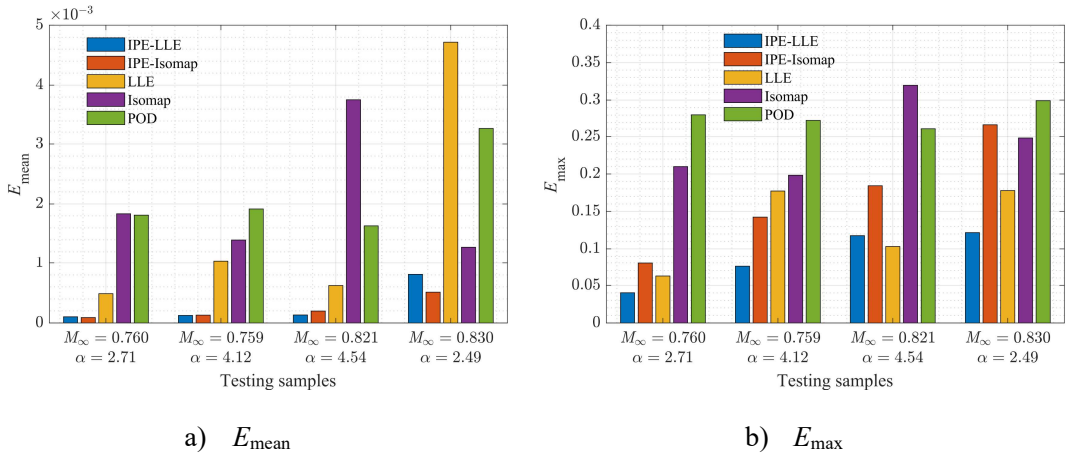

a) $E_{\text{mean}}$      b) $E_{\max}$

Figure 17 The prediction error for different samples under different methods.

The predicted wall pressures of different samples are plotted in Figure 18 as well, where FOM refers to the full-order CFD results. All 5 methods can capture the shock's overall location on the upper and lower surface, but the pressure gradient is slightly smoothed by POD+GPR. Meanwhile, due to the Gibbs phenomenon [41], the pressure distribution ahead of the shock predicted by POD+GPR exhibits small oscillations (Figure 18 a) and b)). However, both IPE-ML and ML+GPR can recover the pressure gradient at the shocks, yet IPE-ML doesn't exhibit non-physical pressure jumps (Figure 18 c)) or shock displacement (Figure 18 d)).

Additionally, due to the global characteristics of POD modes, POD+GPR exhibits large prediction errors at the mid-chord of the lower surface influenced by training samples with shocks existing on the lower surface at higher Mach numbers. Locally-applied IPE-ML and ML+GPR are not impacted, leading to more precise predictions of subsonic regions on lower surface.

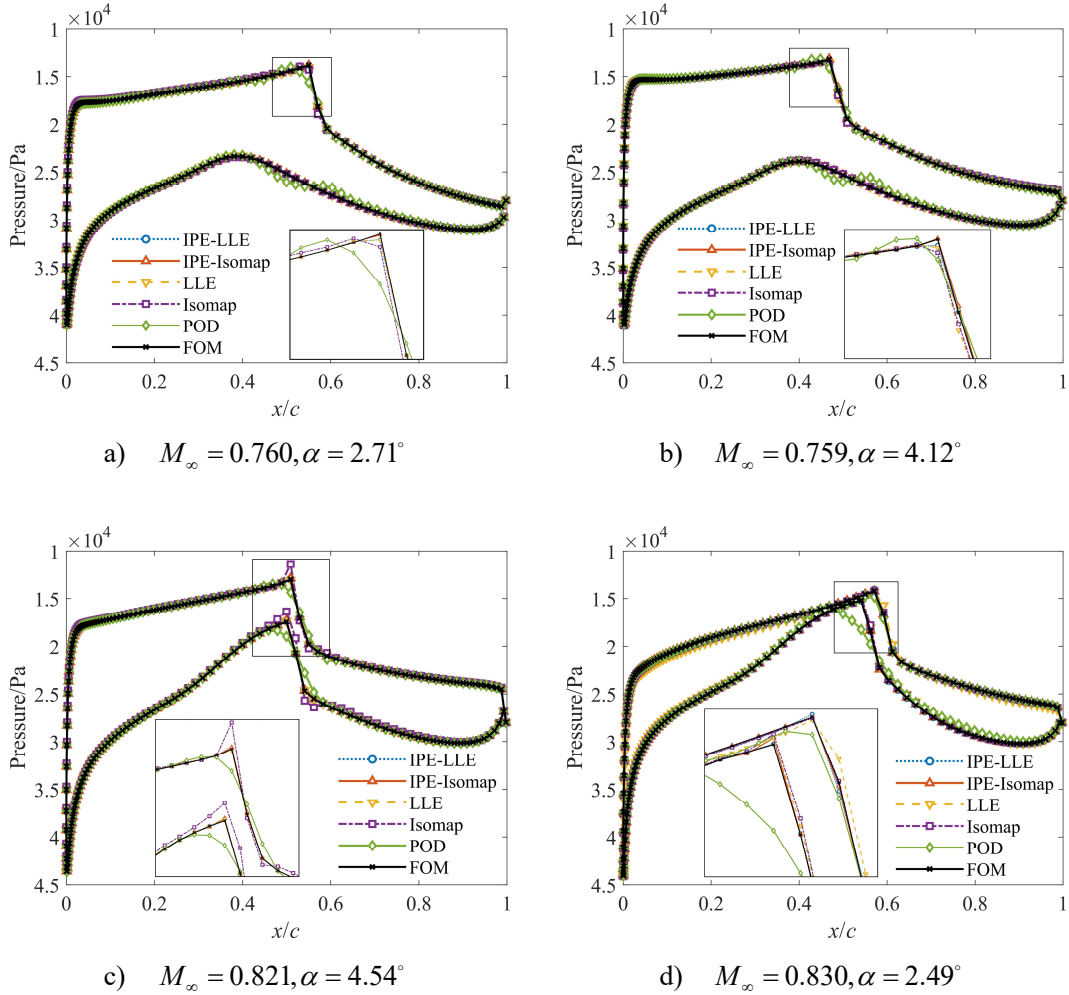

Figure 18 The predictions of wall pressure for 4 testing samples.

We also provide the predictions of full-order flow fields and their corresponding absolute prediction error in Figure 19-Figure 22 and Figure 23-Figure 26 to show the detailed performance in different regions of the entire flow fields. The POD+GPR exhibits more pronounced smoothing of the pressure gradients and significant pressure oscillations, and an anomalous increase in pressure at the mid-chord of the lower surface in Figure 20 f) is also observed. In comparison, IPE+ML and ML+GPR both achieve high prediction accuracy at $M_\infty$=0.759 and 0.760, but their accuracies diverge as $M_\infty$ increases. As $M_\infty$ increases, the local curvature of the manifold changes markedly. Consequently, the manifold coordinates obtained by ML bias more strongly from the true coordinates, leading to errors in the predicted shocks. IPE-ML can further refine these manifold coordinates, improving the flow-field prediction accuracy in the shock regions and confining the dominant errors to the shock center.

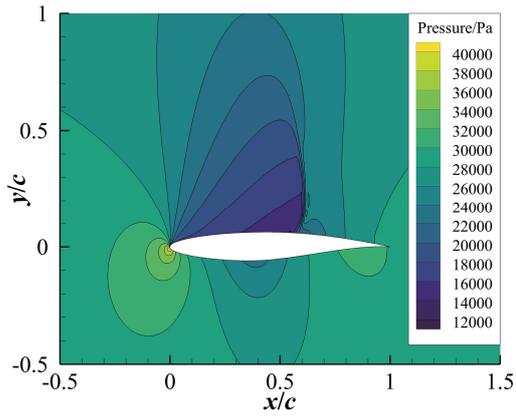

a)  FOM

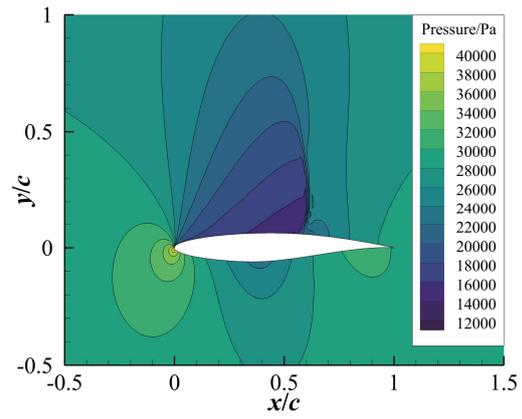

b)  IPE-LLE

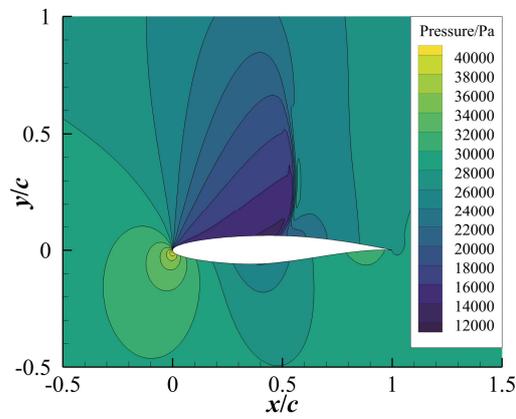

c)  IPE-Isomap

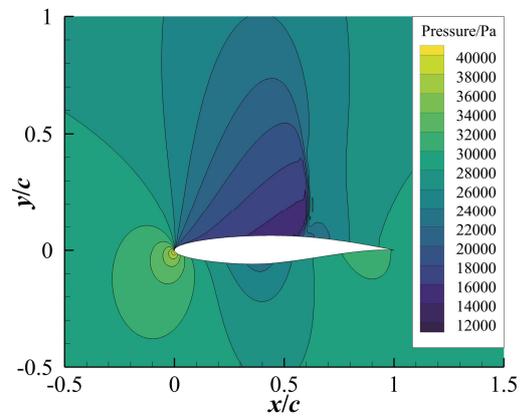

d)  LLE+GPR

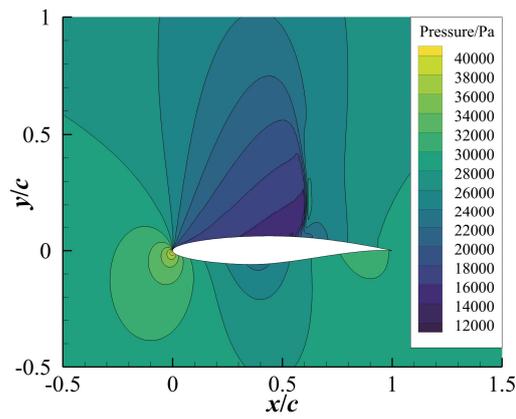

e)  Isomap+GPR

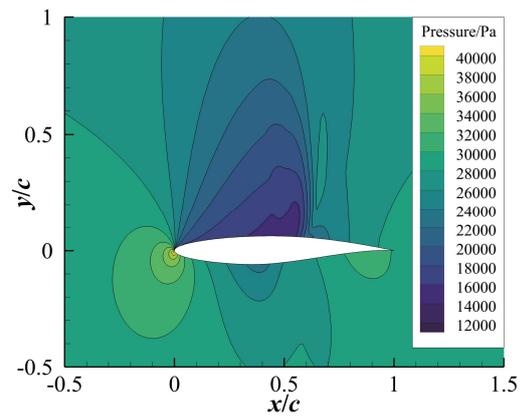

f)  POD+GPR

Figure 19 The predictions of full-order flow field for testing sample at $M_\infty = 0.760, \alpha = 2.71°$.

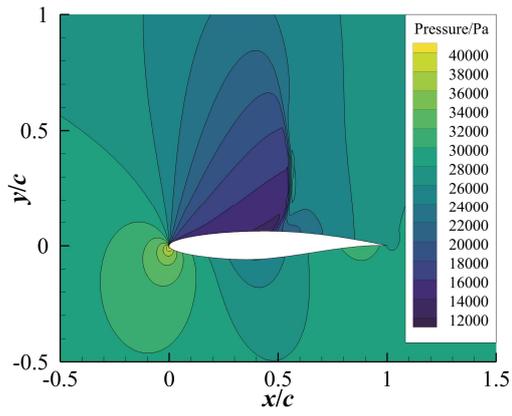
a) FOM

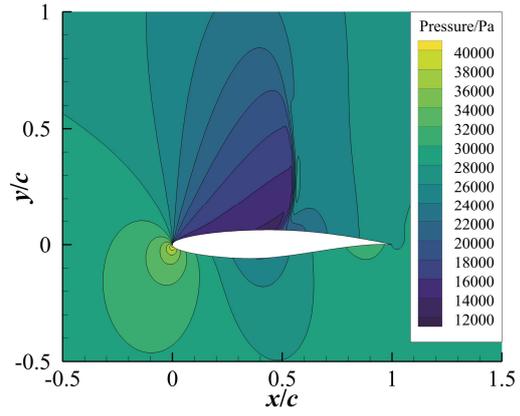
b) IPE-LLE

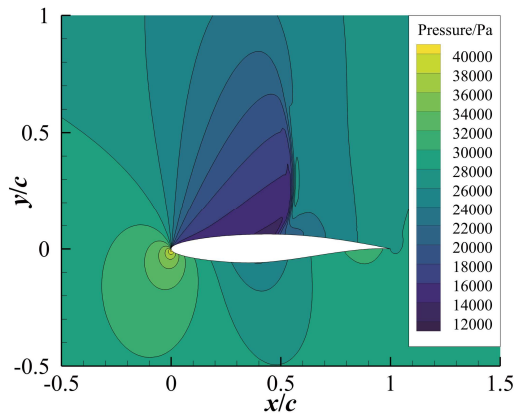
c) IPE-Isomap

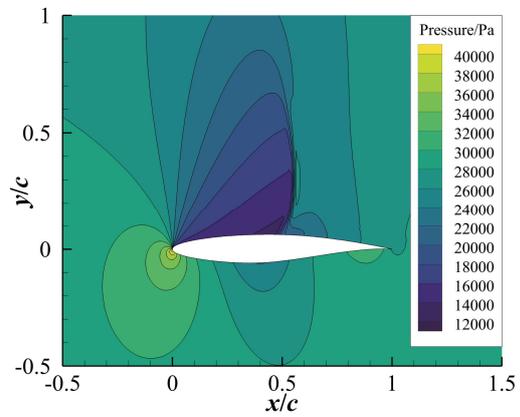
d) LLE+GPR

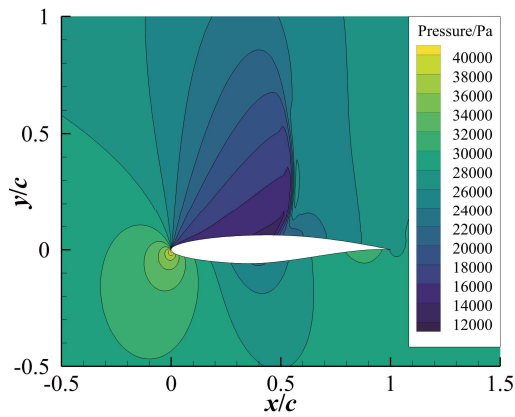
e) Isomap+GPR

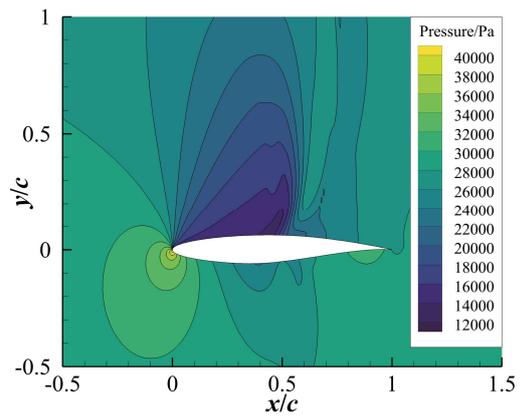
f) POD+GPR

Figure 20 The predictions of full-order flow field for testing sample at $M_\infty = 0.759, \alpha = 4.12°$.

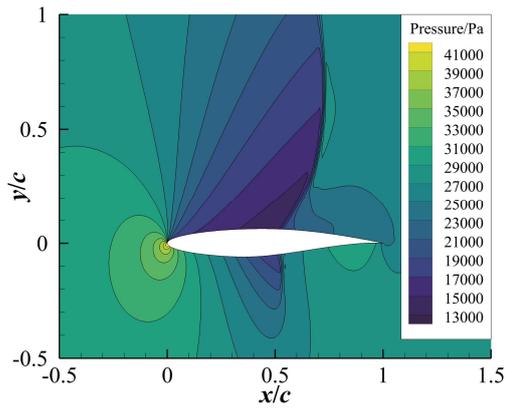

a) FOM

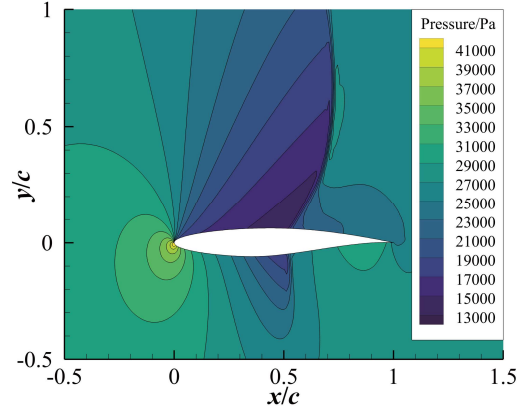

b) IPE-LLE

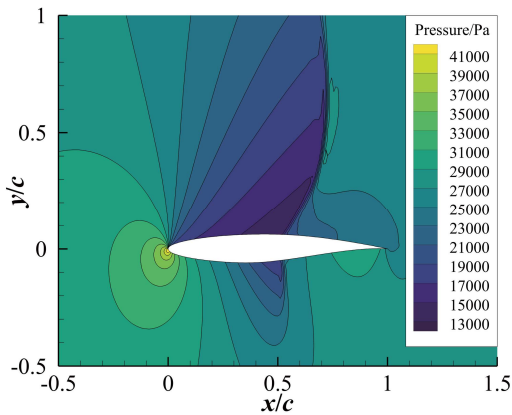

c) IPE-Isomap

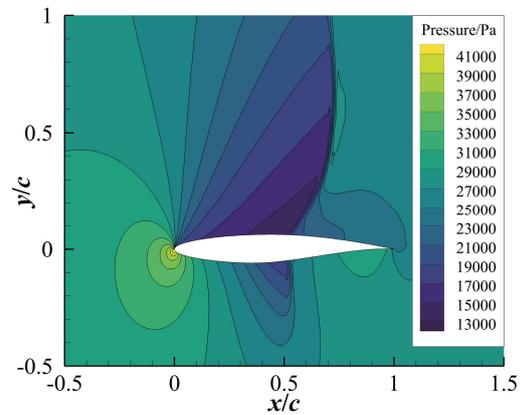

d) LLE+GPR

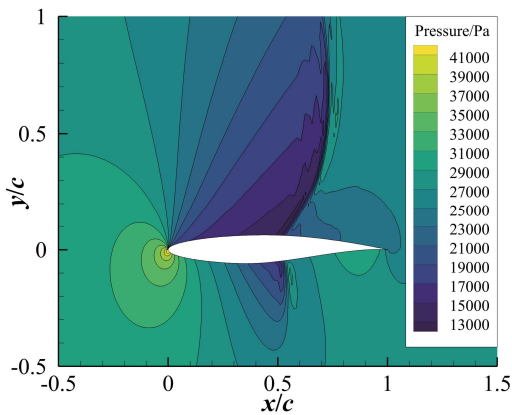

e) Isomap+GPR

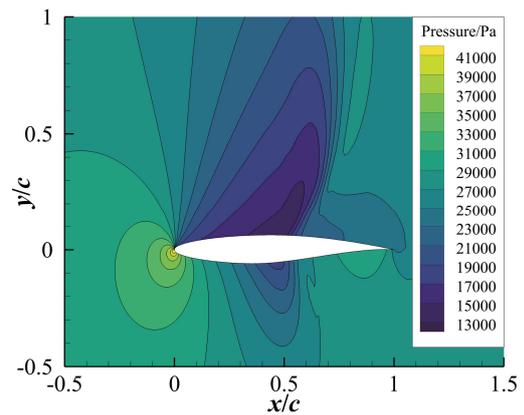

f) POD+GPR

Figure 21 The predictions of full-order flow field for testing sample at $M_\infty = 0.821, \alpha = 4.54°$.

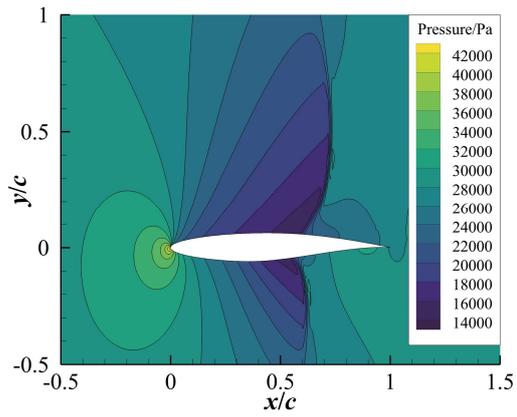

a) FOM

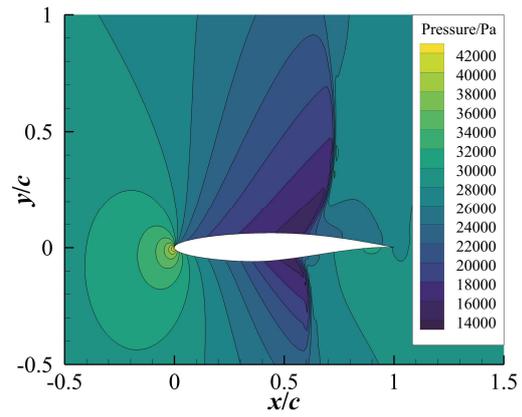

b) IPE-LLE

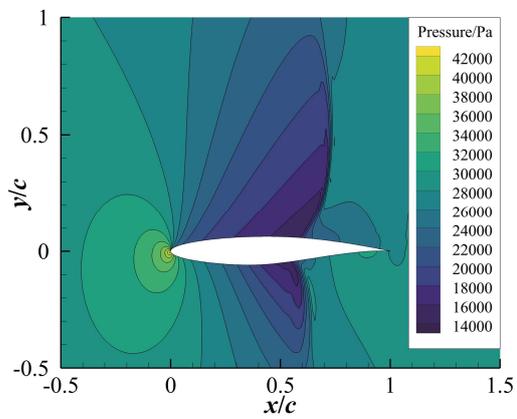

c) IPE-Isomap

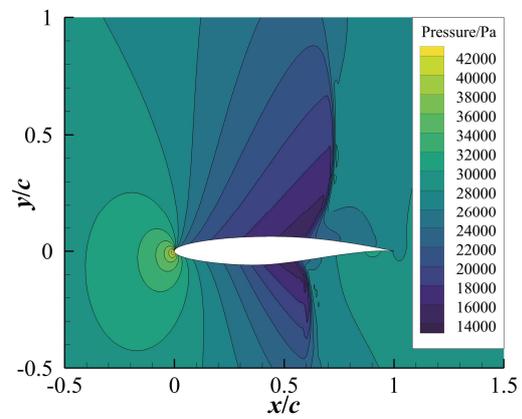

d) LLE+GPR

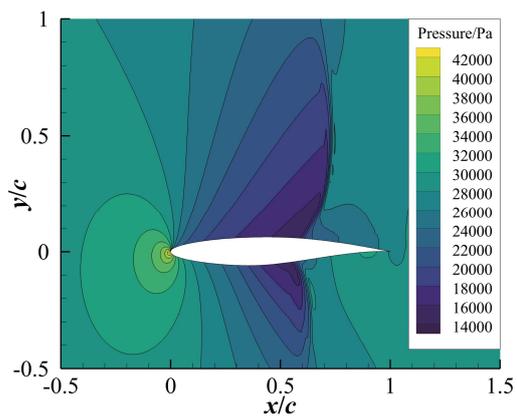

e) Isomap+GPR

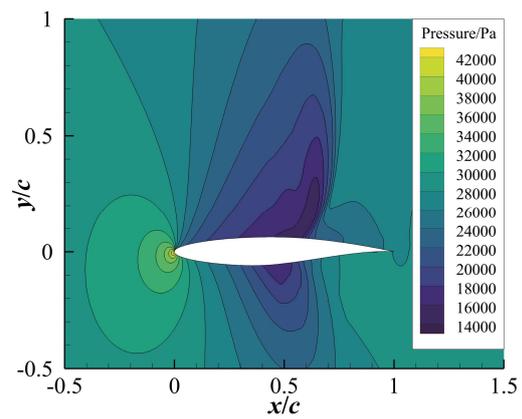

f) POD+GPR

Figure 22 The predictions of full-order flow field for testing sample at $M_\infty = 0.830, \alpha = 2.49°$.

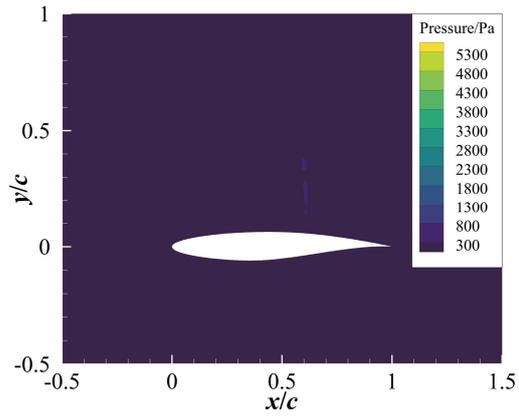 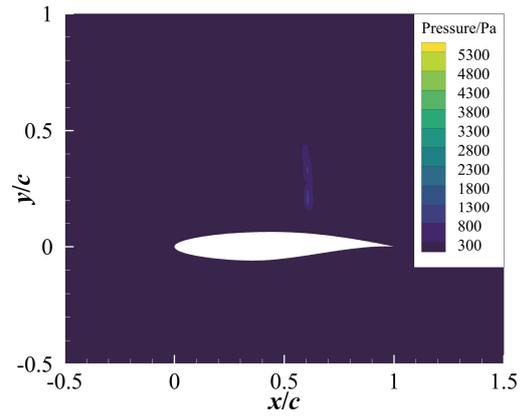

a) IPE-LLE        b) IPE-Isomap

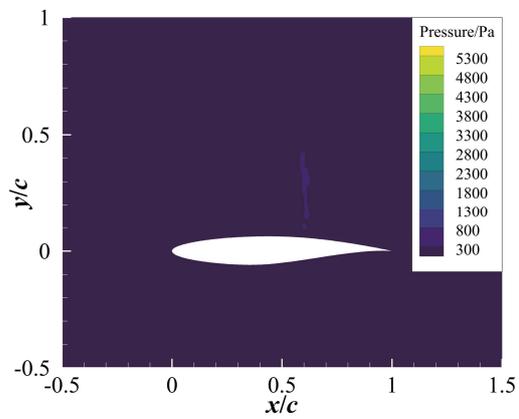 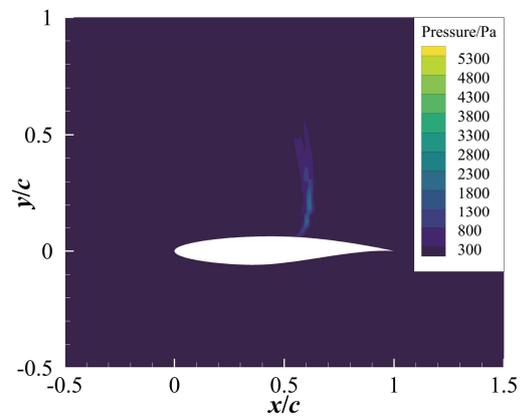

c) LLE+GPR        d) Isomap+GPR

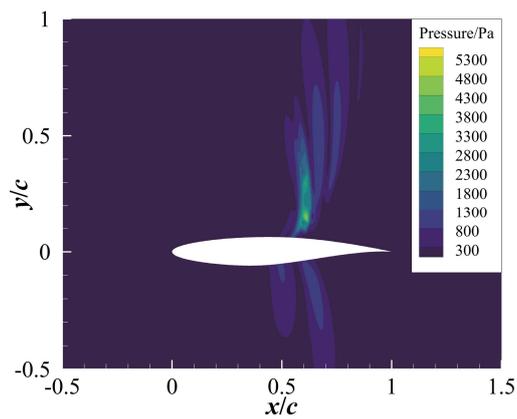

e) POD+GPR

Figure 23 The absolute prediction error of full-order flow field for testing sample at $M_\infty = 0.760, \alpha = 2.71°$.

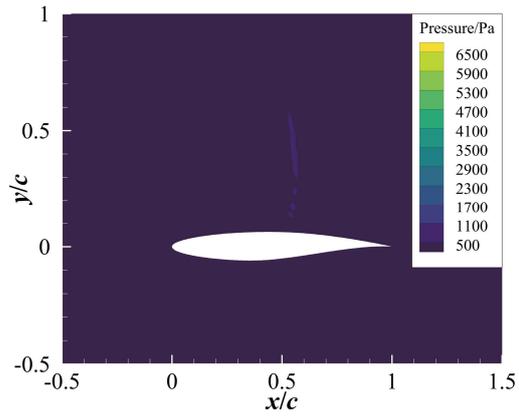
a) IPE-LLE

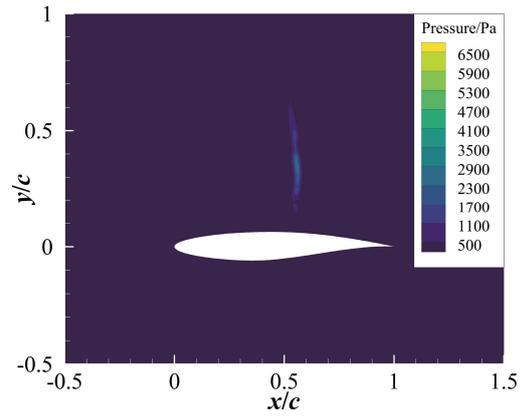
b) IPE-Isomap

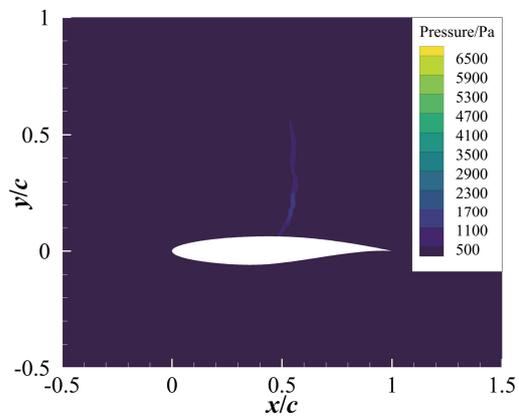
c) LLE+GPR

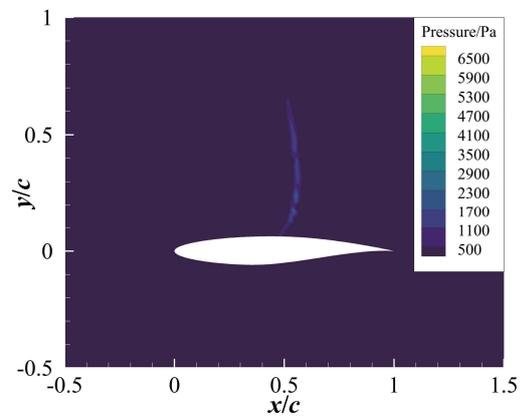
d) Isomap+GPR

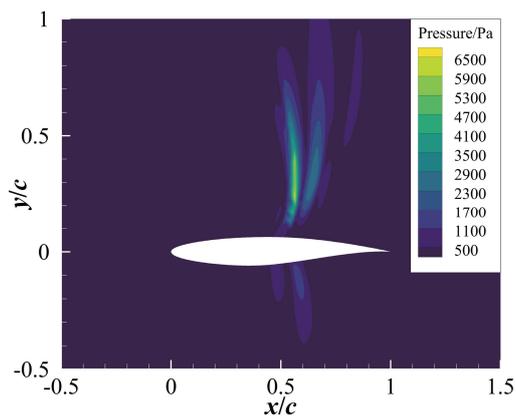
e) POD+GPR

Figure 24 The absolute prediction error of full-order flow field for testing sample at $M_\infty = 0.759, \alpha = 4.12°$.

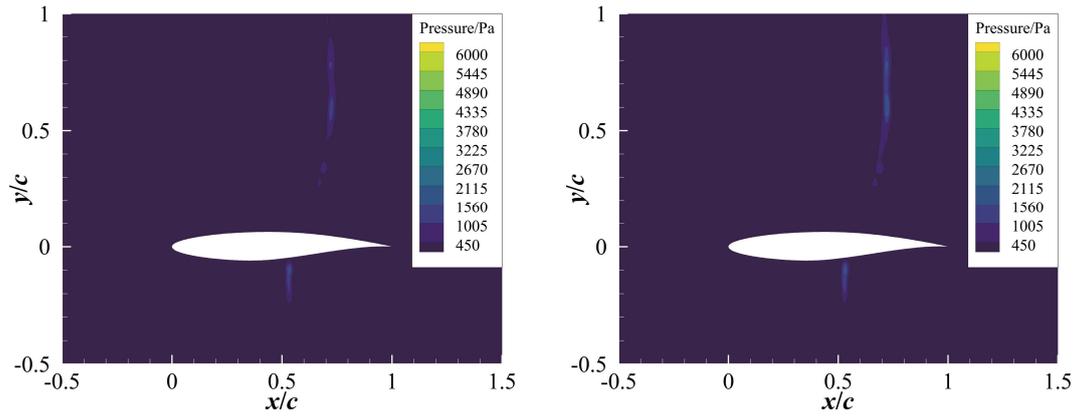

a) IPE-LLE    b) IPE-Isomap

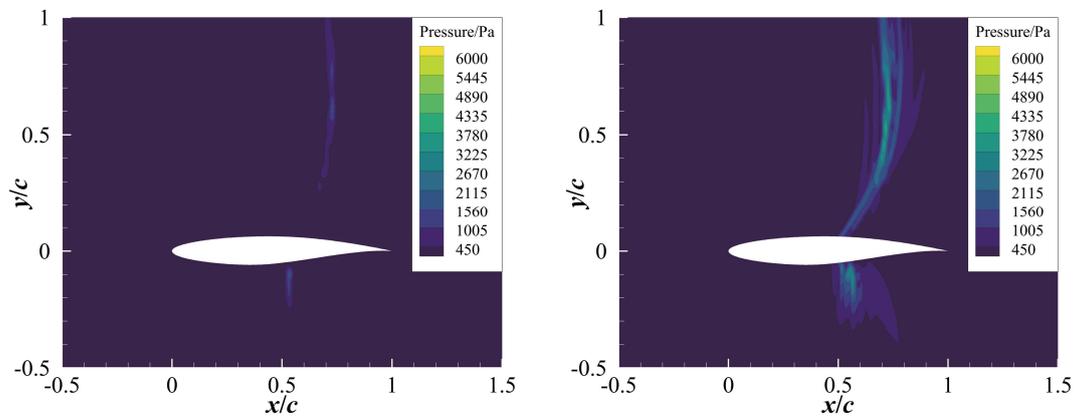

c) LLE+GPR    d) Isomap+GPR

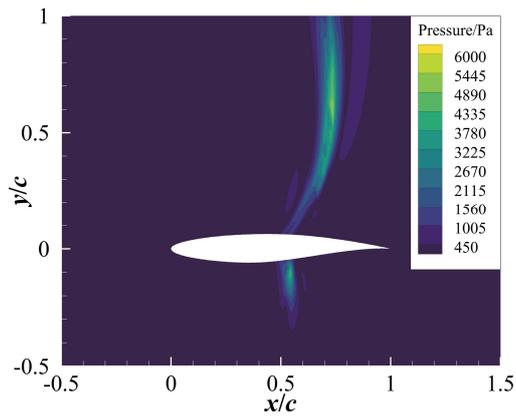

e) POD+GPR

Figure 25 The absolute prediction error of full-order flow field for testing sample at $M_\infty = 0.821, \alpha = 4.54°$.

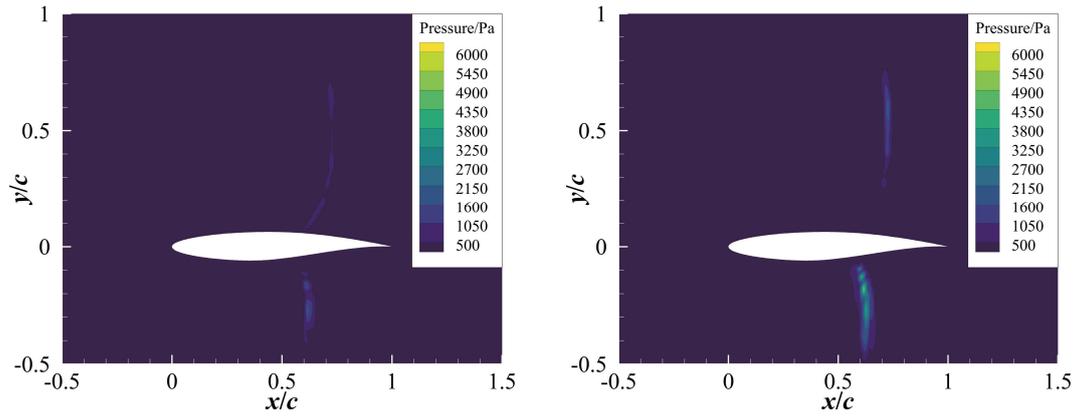

a) IPE-LLE  b) IPE-Isomap

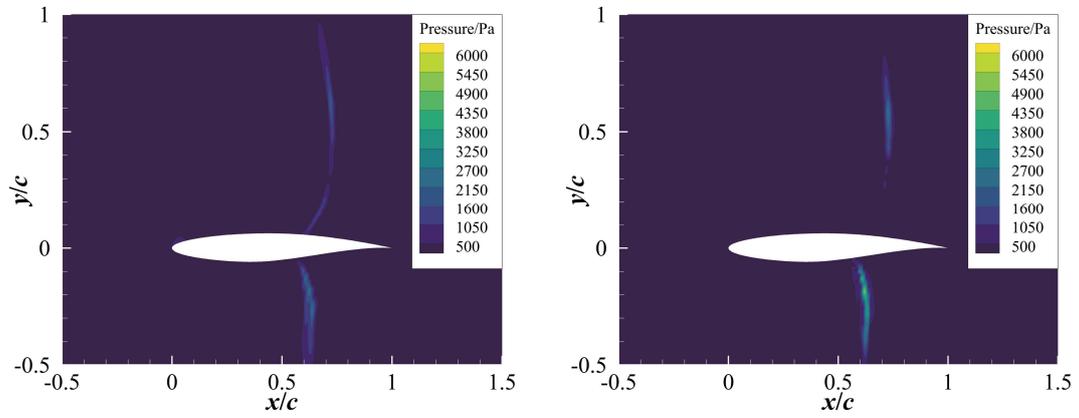

c) LLE+GPR  d) Isomap+GPR

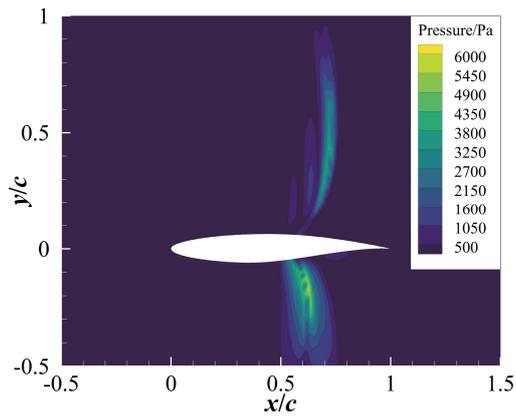

e) POD+GPR

Figure 26 The absolute prediction error of full-order flow field for testing sample at $M_\infty = 0.830, \alpha = 2.49°$.

## B. Test case 2: Supersonic flow fields around hexagon airfoil

1) **Dataset generation**

The second test case is predicting the supersonic flow fields around a hexagon airfoil under $M_\infty \in [1.70, 2.50]$ and $\alpha \in [0°, 4°]$. 550 samples are generated by random LHS as well, in which 500 samples are used as the training set and the rest are included in the testing set (as is shown in Figure 27).

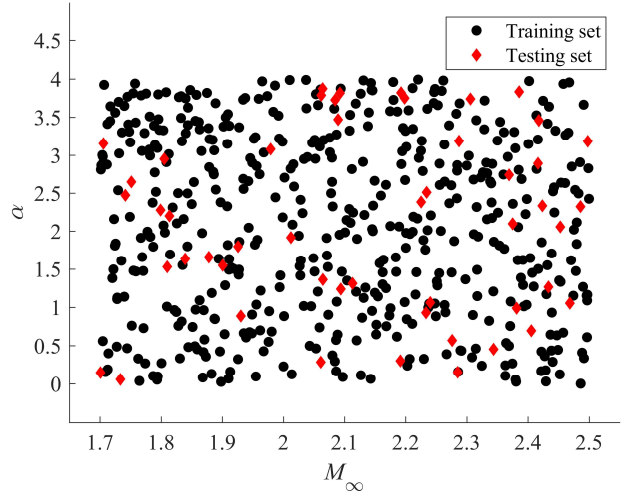

Figure 27 The distribution of supersonic samples in parameter space.

The full-order flow fields are also obtained by RANS-based CFD simulations. A structured grid with y+ < 1 is utilized. The grid convergence study is presented in Appendix A as well. The solver type, turbulence model, boundary condition type and spatial discretization scheme are the same with test case 1. For flux computation, the AUSM scheme is chosen to enhance the stability of CFD solver under supersonic flow.

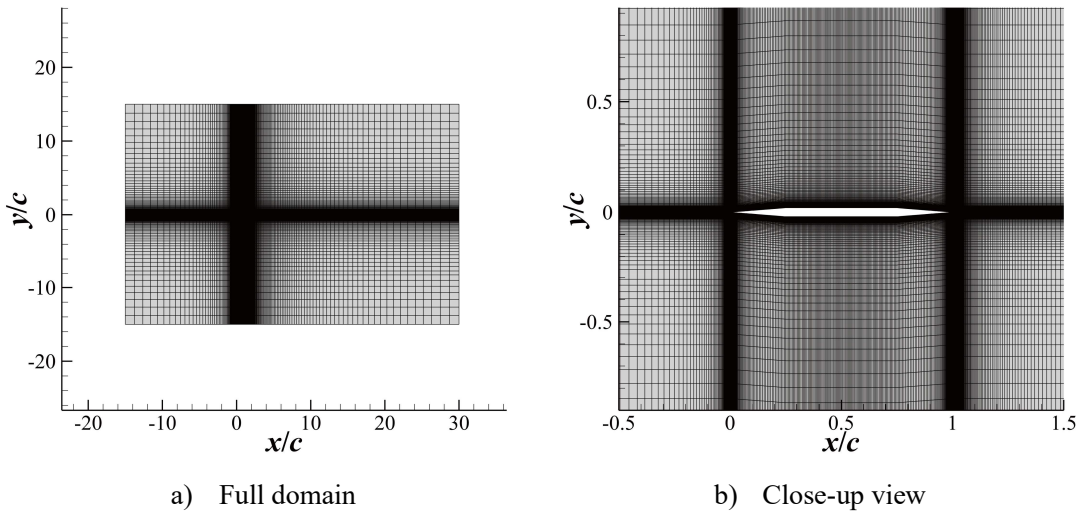

a) Full domain  b) Close-up view

Figure 28 The structured grid for hexagon airfoil (177,608 cells in total).

## 2) Hyperparameter tuning

The hyperparameters in this case are the same as in Table 1, but the values require re-analysis. The L-BFGS settings for IPE-ML remain unchanged. The results of the analysis are as follows.

The $E_{mean}$, $E_{max}$ and $E_{com}$ under different $d$ and $k$ through IPE-LLE and IPE-Isomap are plotted in Figure 29 and Figure 30, separately. For IPE-LLE, $E_{com}$ reaches the smallest at $d = 2$ and $k = 40$, with corresponding $E_{mean}$ and $E_{max}$ of 0.000427 and 0.0353. Meanwhile, for IPE-Isomap, the minimum $E$com occurs at $d = 4$, with $k = 25$, which is different from previous conclusions regarding the dimensionality of the manifold. However, the prediction error of IPE-Isomap at $d = 4$ and $k = 25$ ($E_{mean}$ = 0.000693 and $E_{max}$ = 0.0344) is not significantly different from that at $d = 2$ and $k = 25$ ($E_{mean}$ = 0.000923 and $E_{max}$ = 0.0330). The additional two dimensions are to increase the complexity of the manifold structure to better match the nonlinearity of shock-dominating supersonic flow fields. Therefore, we can still consider that the flow fields are approximately distributed on a 2-dimensional manifold. But in order to achieve higher prediction accuracy, we choose $d = 4$ and $k = 25$ in IPE-Isomap yet.

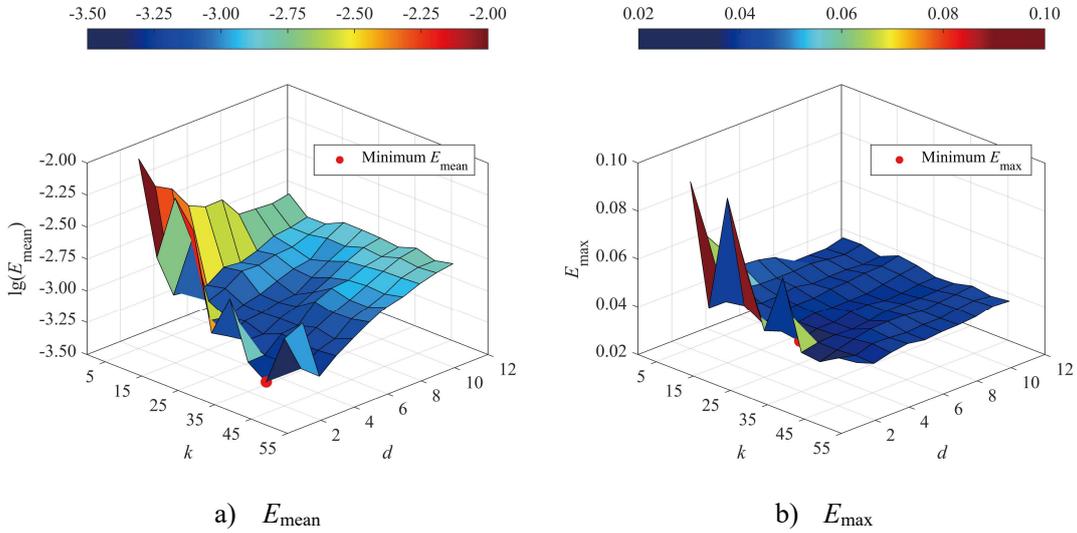

a)  $E_{mean}$        b)  $E_{max}$

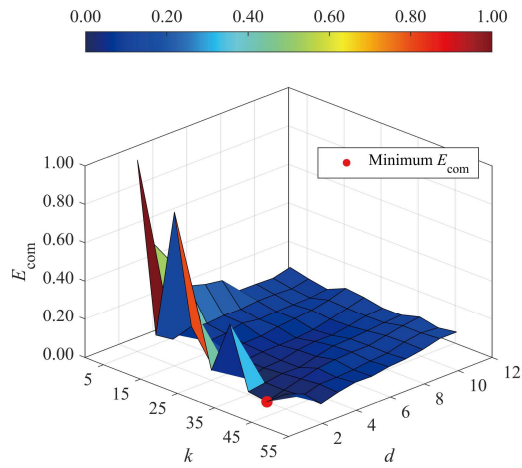

c) $E_{com}$

Figure 29 The prediction error of IPE-LLE under different $d$ and $k$.

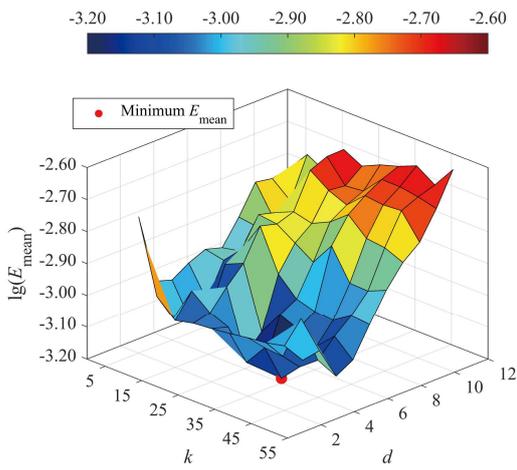

a) $E_{mean}$

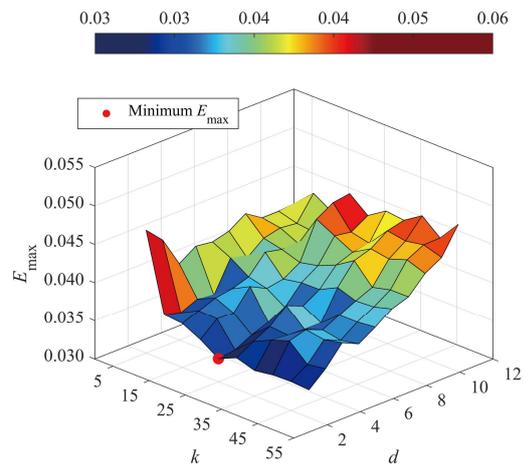

b) $E_{max}$

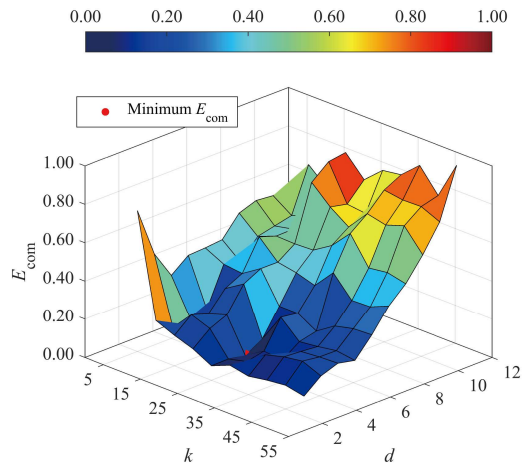

c) $E_{com}$

Figure 30 The prediction error of IPE-Isomap under different $d$ and $k$.

Next, the influence of $v$ and $l$ is shown in Figure 31. The effect of $v$ on the flow-field prediction accuracy is significantly different from that in transonic case. When $v$ is near 1, $E_{mean}$ approaches the smallest, at which the Matérn kernel is smooth and first-order differentiable, making it more beneficial for reconstructing weakly nonlinear regions. However, the smallest $E_{max}$ corresponds to a $v$ around 0.45. Smaller $v$ results in a less smooth kernel, allowing the KRR-DCR model to better capture the sharp changes in strong nonlinearity at shocks, thereby enhancing reconstruction accuracy at these points. From the perspective of $E_{com}$, the improvement in reconstruction accuracy at shocks due to smaller $v$ is more significant for shock-dominating flow fields. Therefore, in the current test case, we set $v = 0.45$.

The best $l$ is around 1.25-1.5, which is of great difference from test case 1 as well. This is because, at supersonic, the locations of shocks do not change significantly, allowing the KRR-DCR model to reconstruct the flow fields using a wider range of modes. Moreover, the fixed positions of shocks also result in higher prediction accuracy for supersonic flow fields compared to transonic flow fields when comparing Figure 8 and Figure 31.

In summary, for the current supersonic test case, we choose $v = 0.45$ and $l = 1.25$.

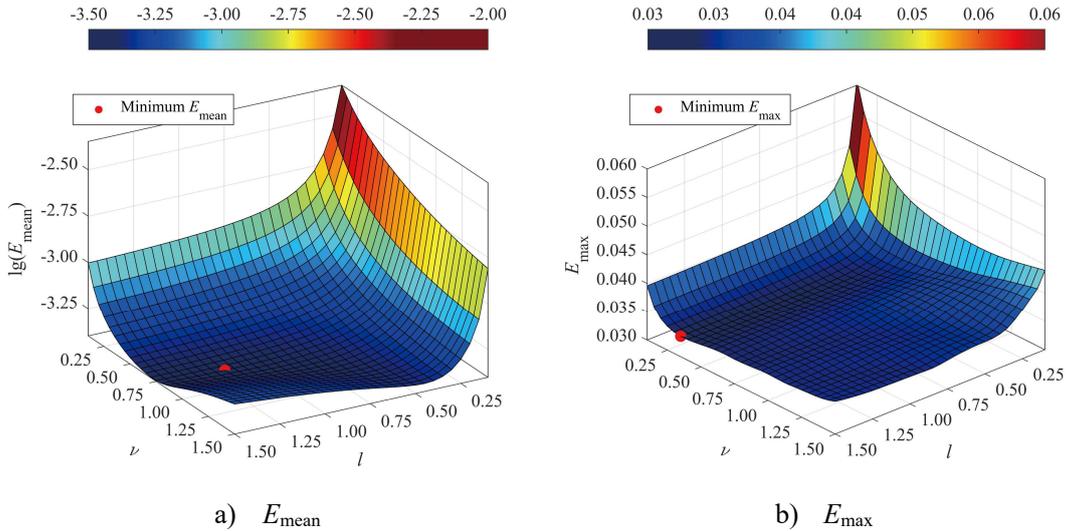

a) $E_{mean}$

b) $E_{max}$

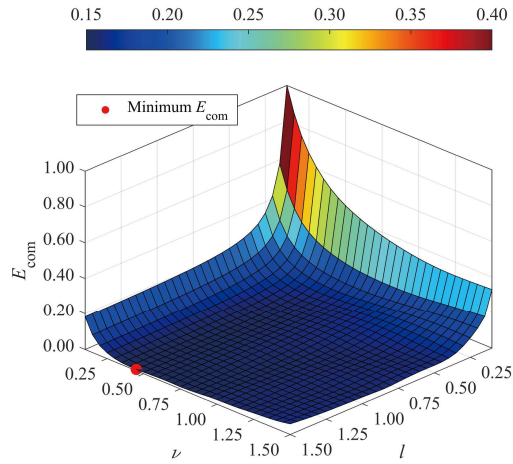

c)  $E_{com}$

Figure 31 The prediction error of IPE-LLE under different $v$ and $l$.

### 3) Flow-field prediction

Similarly, the convergence history of the optimization for IPE-LLE and IPE-Isomap are shown in Figure 32 and Figure 33 Compared with the former case, the supersonic case requires more iterations, for both IPE-LLE and IPE-Isomap, yet both of them converge within 15 iterations.

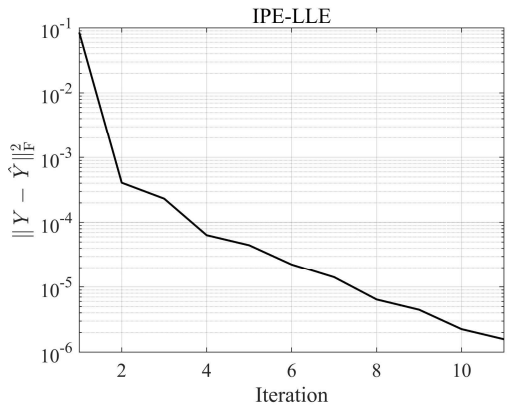

Figure 32 The convergence history for IPE-LLE.

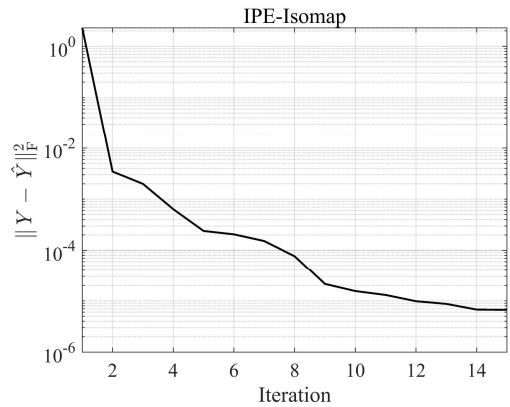

Figure 33 The convergence history for IPE-Isomap.

By comparing the original manifold coordinates with those adjusted using IPE-ML (Figure 34 and Figure 35), it is evident that the manifolds obtained by both IPE-ML are more uniform, without significant clustering of small groups which can be observed on the left side of Figure 34 a).

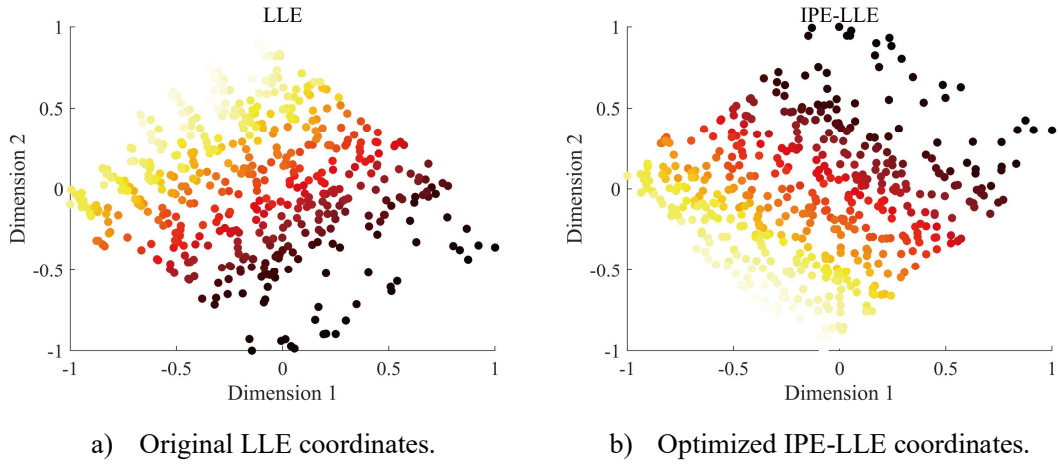

a) Original LLE coordinates.  b) Optimized IPE-LLE coordinates.

Figure 34 Comparison of the original manifold coordinates and optimized manifold coordinates for IPE-LLE.

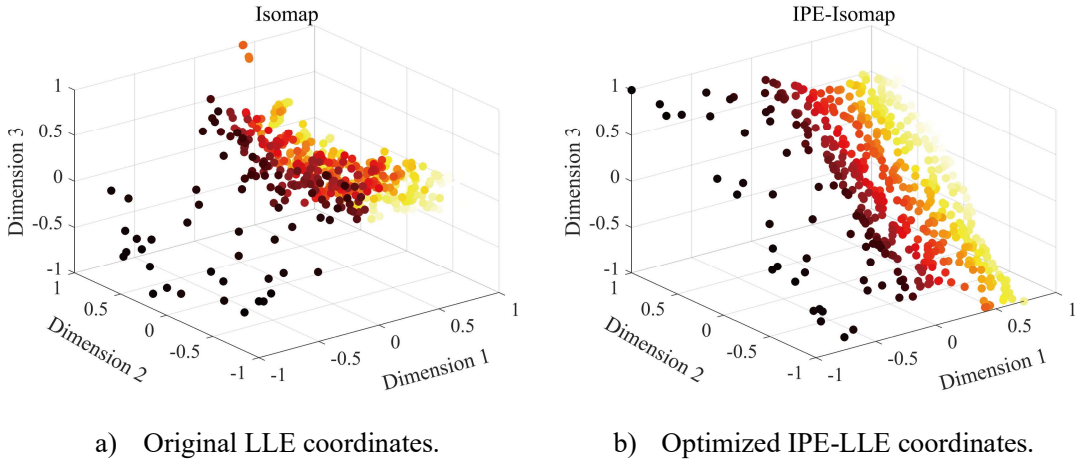

a) Original LLE coordinates.  b) Optimized IPE-LLE coordinates.

Figure 35 Comparison of the original manifold coordinates and optimized manifold coordinates for IPE-Isomap (the first 3 dimensions).

Then, the prediction error of 5 methods are compared in Figure 36. In this case, $E_{mean}$ of LLE+GPR and Isomap+GPR are even much higher than that of POD+GPR. Although $E_{mean}$ for IPE-LLE and IPE-Isomap are reduced by about half compared to LLE+GPR and Isomap+GPR methods, there is no significant difference from POD+GPR. Meanwhile, $E_{max}$ for 5 methods are all concentrated within 3%-5%.

To further analyze the reasons, we also randomly select 4 testing samples and calculate their flow-field prediction errors in Figure 37. Among the four methods, LLE+GPR and Isomap+GPR exhibit poorer generalization capability and both show outliers (as shown in Figure 37 a)). However, after embedding physical information through IPE-ML, the generalization capability is significantly improved. Meanwhile, in Figure 37b), $E_{max}$ for each testing sample across different methods do not differ much, which is the same with the observations in Figure 36 b).

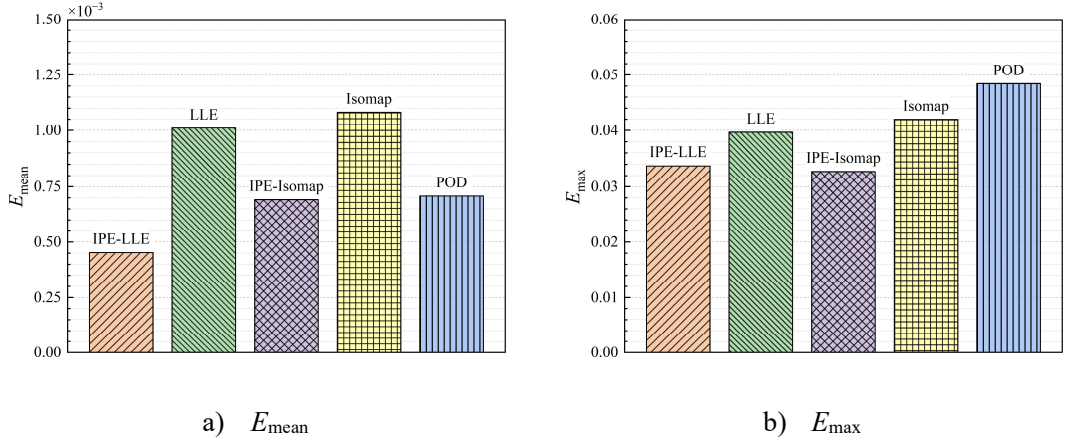

a)  $E_{\text{mean}}$     b)  $E_{\text{max}}$

Figure 36 $E_{\text{mean}}$ and $E_{\text{max}}$ on the testing set for the different methods.

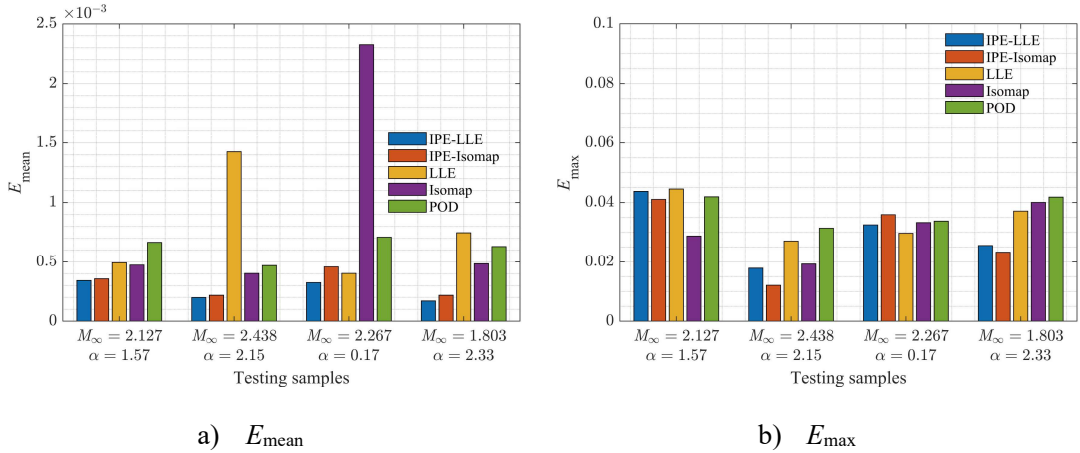

a)  $E_{\text{mean}}$     b)  $E_{\text{max}}$

Figure 37 The prediction error for different samples under different methods.

Since the features of supersonic flow fields are quite similar, we only plot the predicted flow fields and absolute error contours for the testing samples with the largest and smallest $E_{\text{max}}$ ( $M_\infty = 2.127, \alpha = 1.57°$ and $M_\infty = 2.438, \alpha = 2.15°$ ), as shown in Figure 38 and Figure 39. Since the location and number of shocks are relatively fixed under the current supersonic regime, all methods capture the shock positions accurately.

However, from the absolute errors of the two samples, it is evident that the prediction errors of IPE-LLE and IPE-Isomap are concentrated in a small region near the shocks on the trailing edge of the upper surface instead of being spread across the entire domain, just as what POD+GPR gets.

The results above indicate that though IPE-ML cannot significantly reduce the maximum flow-field prediction error, it can control the regions where errors occur and improves the overall prediction accuracy.

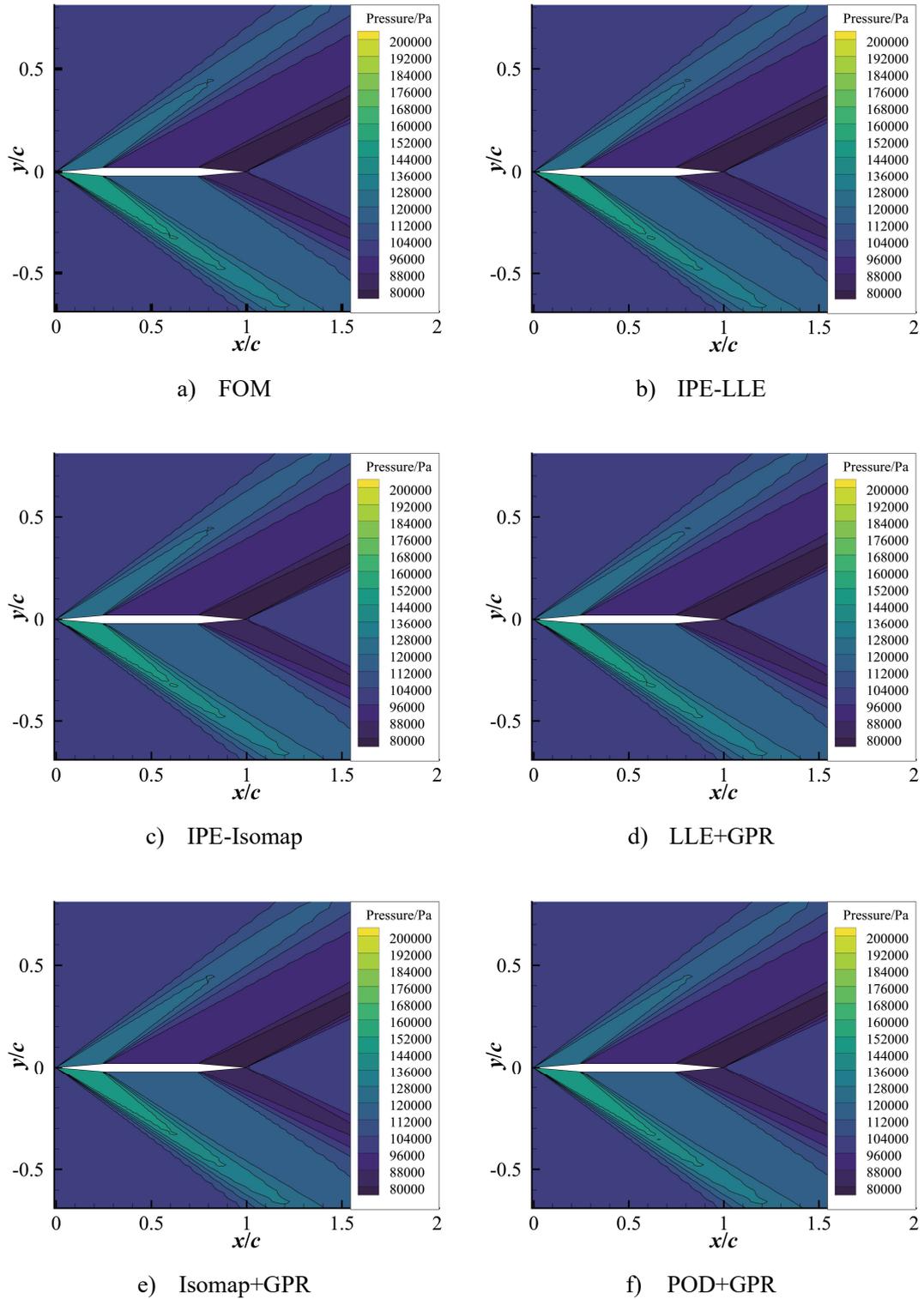

Figure 38 The predictions of full-order flow field for testing sample with largest $E_{max}$ at $M_\infty = 2.127, \alpha = 1.57°$.

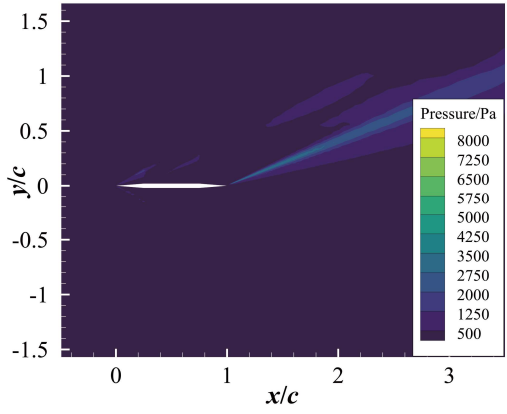

a) IPE-LLE

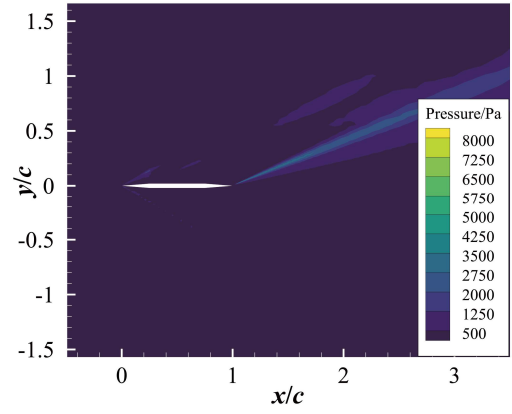

b) IPE-Isomap

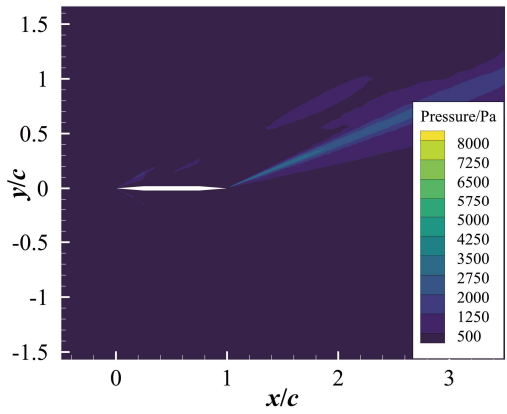

c) LLE+GPR

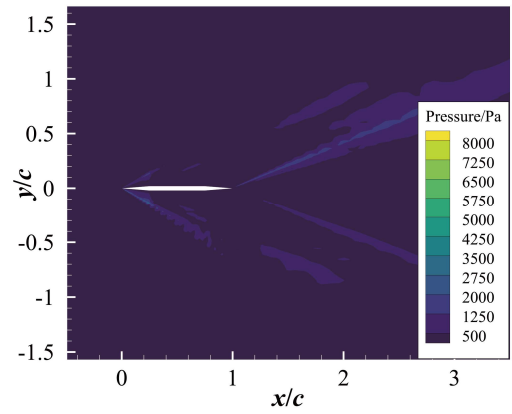

d) Isomap+GPR

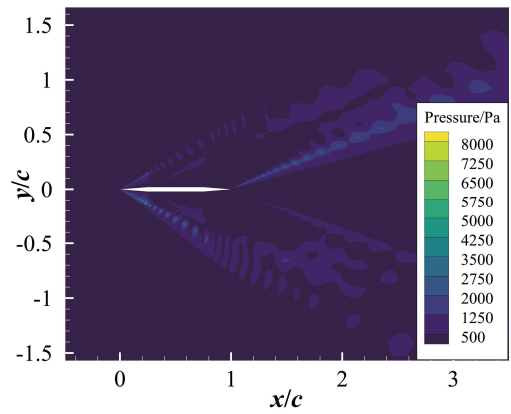

e) POD+GPR

Figure 39 The absolute prediction error of full-order flow field for testing sample with largest $E_{max}$ at $M_\infty = 2.127, \alpha = 1.57°$.

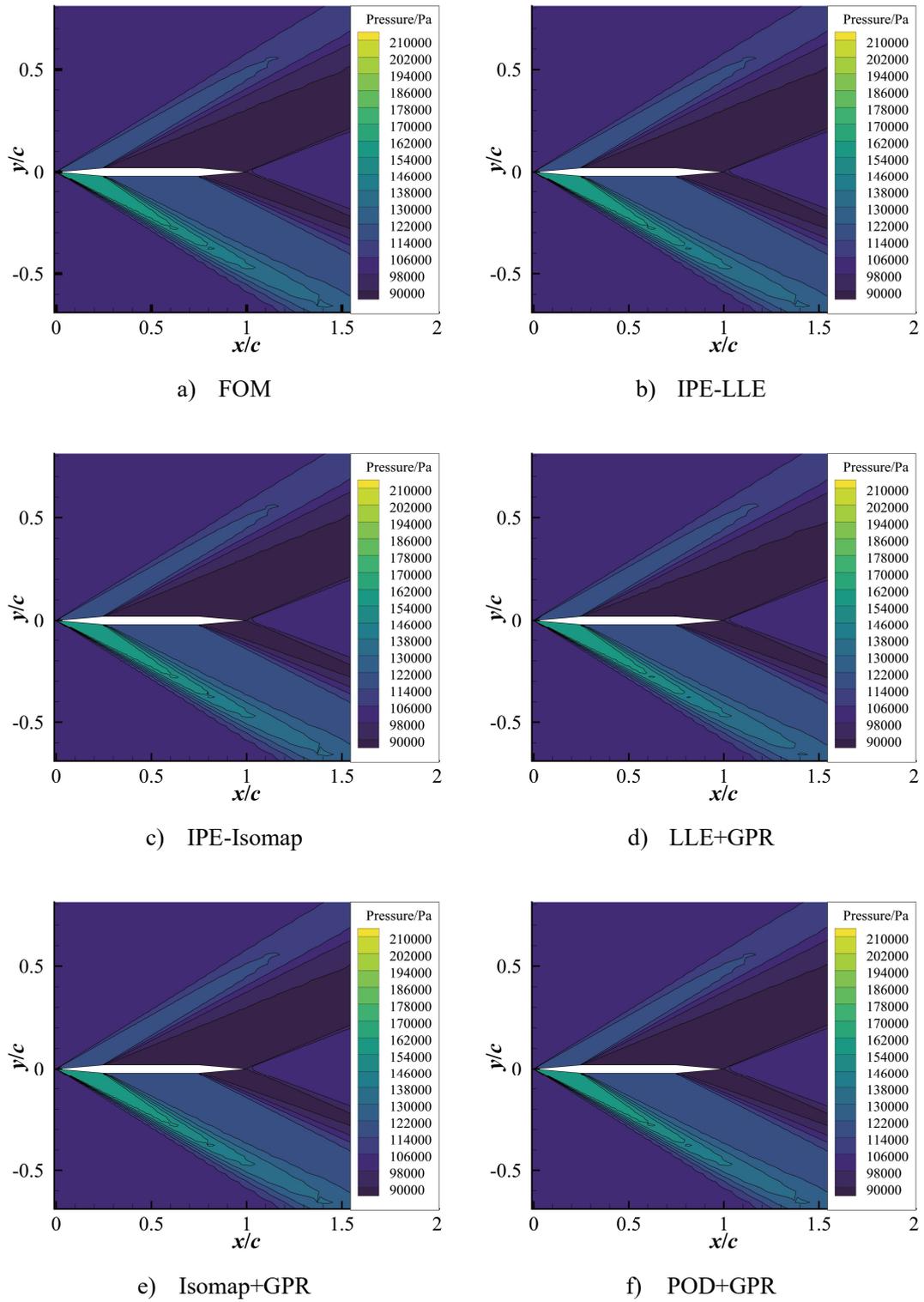

Figure 40 The predictions of full-order flow field for testing sample with the smallest $E_{\max}$ at $M_\infty = 2.438, \alpha = 2.15°$.

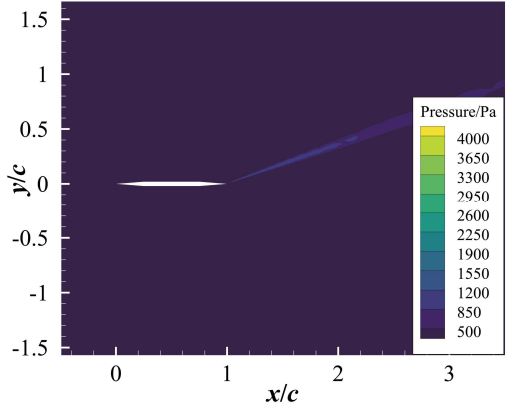

a) IPE-LLE

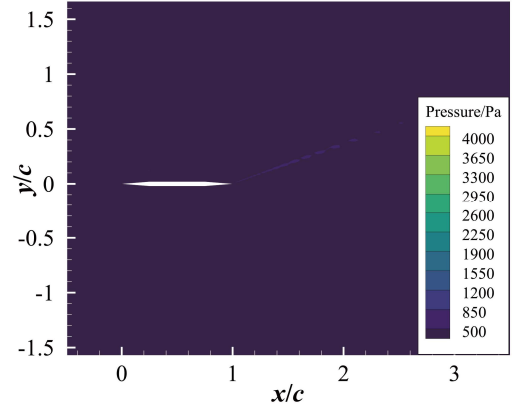

b) IPE-Isomap

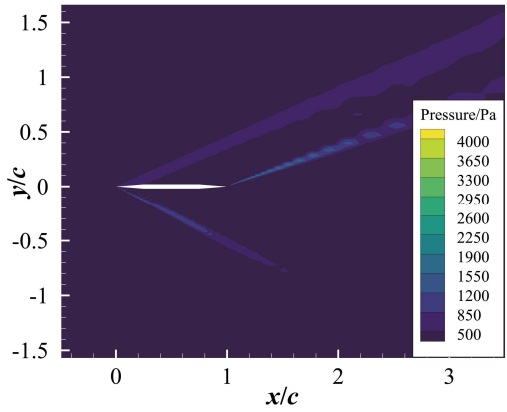

c) LLE+GPR

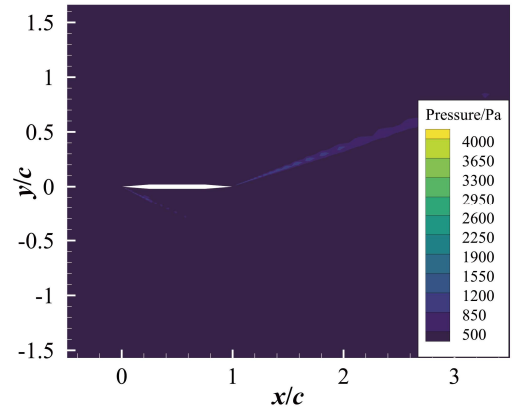

d) Isomap+GPR

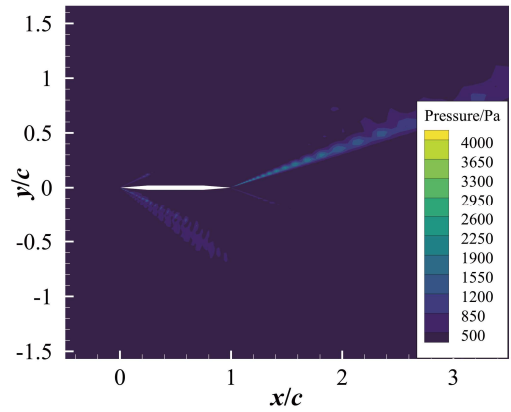

e) POD+GPR

Figure 41 The absolute prediction error of full-order flow field for testing sample with the smallest $E_{\max}$ at $M_\infty = 2.438, \alpha = 2.15°$.

## C. Time consumption

The time consumption for modeling and flow-field prediction by different methods is compared in Table 3 and Table 4. All time shown in these table is an average over 20 repeated experiments. Although the proposed IPE-ML introduces an optimization during dimensionality reduction, the additional time consumption is acceptable because the optimization converges rapidly. During the surrogate-model training and flow-field prediction phases, the IPE-ML introduces no additional cost compared to the conventional ML+GPR methods.

Table 3 Time consumption for transonic case ($s^{-1}$)

| Stage \ Method | LLE+GPR | Isomap+GPR | IPE-LLE | IPE-Isomap | POD |
|---|---|---|---|---|---|
| Dimensionality reduction | 1.6358 | 0.5865 | 2.3364 | 1.5718 | 0.9578 |
| Surrogate-model training | 0.0049 | 0.0053 | 0.0084 | 0.0084 | 0.0268 |
| Flow-field prediction | 1.0744 | 1.1272 | 1.0954 | 1.1150 | 0.1204 |

Table 4 Time consumption for supersonic case ($s^{-1}$)

| Stage \ Method | LLE+GPR | Isomap+GPR | IPE-LLE | IPE-Isomap | POD |
|---|---|---|---|---|---|
| Dimensionality reduction | 9.9813 | 1.1963 | 10.671 | 2.4932 | 2.2914 |
| Surrogate-model training | 0.0052 | 0.0046 | 0.0104 | 0.0093 | 0.0261 |
| Flow-field prediction | 1.6070 | 1.7748 | 1.6058 | 1.6043 | 0.1824 |

## 5. Conclusions

In this paper, a novel nonlinear reduced-order model with implicit physics embedding IPE-ML is proposed to strength the flow-field prediction accuracy of the conventional ML methods. The main conclusions are drawn as follows:

1. Based on conventional ML methods (Isomap and LLE), we introduce manifold-coordinate optimization to embed physical parameters such as angle of attack and Mach number into the low-dimensional coordinate system. Compared with POD + GPR and ML + GPR, the proposed method reduces the overall prediction error for transonic and supersonic flow fields, and also improves prediction accuracy in highly nonlinear shock regions.
2. Lipschitz continuity analysis and error decomposition theoretically prove that under appropriately chosen sample size and hyperparameters, the proposed IPE-ML can further improve flow-field prediction accuracy on the basis of conventional ML+GPR.
3. After adjustment with IPE-ML, the global manifold geometry remains unchanged,

undergoing only local fine-tuning. Moreover, IPE-ML effectively fixes disconnected neighborhood graph that arises during dimensionality reduction.
4. All manifold-coordinate optimizations converge within 15 iterations which make the additional time consumption acceptable.
5. Future work includes addressing variable boundary conditions and unsteady cases, and extending the method to reduced-order modeling and physical embedding of flow-control equations.

**Appendix A**

The grid convergence studies for 2 test cases are listed below. The simulation condition of two studies are $M_\infty = 0.734$, $\alpha = 2.79°$, $Re = 6.5 \times 10^5$ and $M_\infty = 2.00$, $\alpha = 2.00°$, $Re = 3.0 \times 10^6$, respectively. And the CFD solver settings are identical to those described in section 3.

For test case 1 (transonic regime), the numerical results for $C_l$ and $C_d$ obtained on the fine mesh are essentially indistinguishable from those on the extra-fine mesh, and the grid convergence index (GCI) is markedly reduced. Accordingly, the fine mesh is adopted for sample generation in this case.

For test case 2 (supersonic regime), both the medium and fine meshes yield results close to the extra-fine mesh, with only a modest GCI improvement. Nevertheless, because supersonic flows demand accurate shock capturing, the fine mesh with the finer spatial discretization is selected.

Table 5 Grid convergence study of test case 1

| Grid level | Total cells | $C_l$ | $C_d$ | $GCI_{C_l}$ /% | $GCI_{C_d}$ /% |
|---|---|---|---|---|---|
| Coarse | 28,710 | 0.7666 | 0.1709 | 2.092 | 2.324 |
| Medium | 49,020 | 0.7757 | 0.1735 | 1.131 | 1.339 |
| Fine | 82,844 | 0.7806 | 0.1750 | 0.344 | 0.267 |
| Extra-Fine | 140,456 | 0.7821 | 0.1753 | | |

Table 6 Grid convergence study of test case 2

| Grid level | Total cells | $C_l$ | $C_d$ | $GCI_{C_l}$ /% | $GCI_{C_d}$ /% |
|---|---|---|---|---|---|
| Coarse | 36,696 | 0.08259 | 0.01533 | 2.681 | 0.325 |
| Medium | 80,982 | 0.08200 | 0.01531 | 1.110 | 0.327 |
| Fine | 177,608 | 0.08176 | 0.01529 | 0.713 | 0.159 |

| | Extra-Fine | 401,856 | 0.08160 | 0.01528 |

## Appendix B

In POD, the flow fields are decomposed into a set of orthogonal modes, ranked by their associated eigenvalues. Each eigenvalue corresponds to the amount of energy in the flow fields that the corresponding POD mode captures. So, we usually compute the cumulative energy ratio $E(r)$ to choose the optimal number of POD modes

$$E(r) = \frac{\sum_{j=1}^{r} \lambda_j}{\sum_{j=1}^{N} \lambda_j} \quad (33)$$

where $r$ is the number of POD modes; $N$ is the total number of eigenvalues; $\lambda_j$ is the $j$th eigenvalue sorted in descending order.

To ensure that POD modes fully capture the characteristics of the flow fields, we aim for $E(r) \approx 1$ [42]. However, due to the presence of noise in the flow fields, in practical applications, a threshold (such as $E(r)=0.99$) is typically used as a criterion to select the number of POD modes.

$E(r)$ of the transonic case and the supersonic case under different $r$ are shown in Figure 42 and Figure 43. When $E(r)$ reaches 0.99, the corresponding $r$ are 12 and 26, separately.

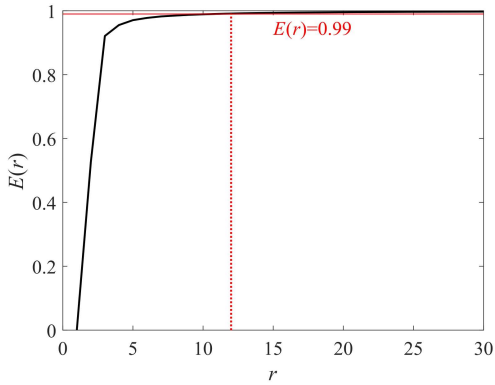 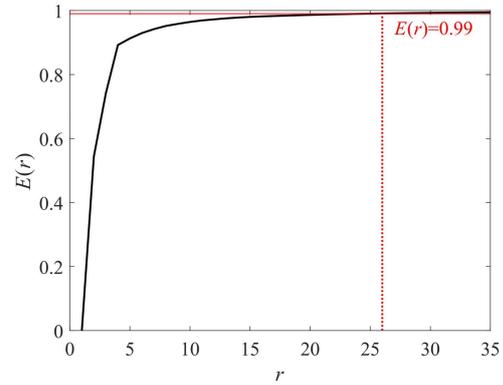

Figure 42 $E(r)$ of the transonic case.  Figure 43 $E(r)$ of the supersonic case.